\documentclass[sort,review]{elsarticle}

\pdfoutput=1

\usepackage{lineno}
\usepackage[colorlinks = true, linkcolor = blue, urlcolor = blue, citecolor = blue, anchorcolor = blue]{hyperref}
\modulolinenumbers[5]
\usepackage{todonotes}
\usepackage{pgfplots} 
\usepackage{graphicx}
\usepackage{booktabs}
\usepackage{geometry}
\usepackage[normalem]{ulem}
\useunder{\uline}{\ul}{}
\usepackage{longtable}
\usepackage{bchart}
\usepackage{pifont}
\usepackage{float}
\usepackage{caption}
\usepackage{subcaption}
\usepackage{comment}
\usepackage{setspace}

\usepackage{booktabs}
\usepackage{pdfpages}

\usepackage{ifthen}
\usepackage{amssymb}
\usepackage[normalem]{ulem} 
\newboolean{showcomments}
\setboolean{showcomments}{true} 
\ifthenelse{\boolean{showcomments}}
  {
		\newcommand{\nbb}[2]{
		\fcolorbox{black}{yellow}{\bfseries\scriptsize#1}
		{$\blacktriangleright$\textcolor{blue}{\textit{#2}}$\blacktriangleleft$}
		}
		
		\newcommand{\remarks}[1]{\color{red}[#1]\color{black}}

		\newcommand{\del}[1]{\textcolor{red}{\sout{#1}}} 
		
		\newcommand{\removed}[1]{\cbstart\removedfragile{#1}\cbend{}}
		\newcommand{\removedfragile}[1]{{\color{red}{\sout{#1}}}{}}

  }
  {
		\newcommand{\nbb}[2]{}
		\newcommand{\remarks}[1]{}

		\newcommand{\del}[1]{} 
		
		\newcommand{\removed}[1]{} 
  		\newcommand{\removedfragile}[1]{}

  }

\journal{International Journal of Human Computer Studies}

\biboptions{authoryear}

\begin{document}

\begin{frontmatter}
\title{Mapping aids using source location tracking increase novices' performance in programming cyber-physical systems}

\author[label1]{Thomas Witte\corref{contribution}}
\ead{thomas.witte@uni-ulm.de}
\author[label2]{Andrea Vogt\corref{contribution}}
\ead{andrea.vogt@uni-ulm.de}
\cortext[contribution]{Authors contributed equally}
\author[label2]{Tina Seufert}
\ead{tina.seufert@uni-ulm.de}
\author[label1]{Matthias Tichy}
\ead{matthias.tichy@uni-ulm.de}

\address[label1]{
    Institute of Software Engineering and Programming Languages \\
    Ulm University, Germany \\
}
\address[label2]{
    Department Learning and Instruction \\Institute of Psychology and Education\\ Ulm University, Germany 
}

\begin{abstract}
\paragraph{Motivation and Background}
Novices need to overcome initial barriers while programming cyber-physical systems behavior, like coding quadcopter missions, and should thus be supported by an adequately designed programming environment. Using multiple representations by including graphical previews is a common approach to ease coding and program understanding. However, novices struggle to map information of the code and graphical previews. Previous studies imply that mapping aids in a live programming environment might support novices while programming and foster a deeper understanding of the content. To implement these mapping aids in a domain independent way Source Location Tracking  based on run-time information can be used.

\paragraph{Method} In our study, we tested $N=82$ participants while interacting and learning in an online programming environment. Using our 2x2 between-subject design study, we investigated the effects of two mapping aids: highlighting and dynamic linking on coding correctness including typical errors, and learning outcomes. Based on process data successful strategies were analyzed. 

\paragraph{Results} 
Combining both mapping aids compared to one aid resulted in higher performance. While highlights were more helpful for implementing the quadcopter missions (mission 1: $p=.008$*, $\eta^2_{\footnotesize\textit{partial}}=.091$), dynamic linking improved learning outcomes on the comprehension ($F(1,75)=5.61$, $p=.020$*, $\eta^2_{\footnotesize\textit{partial}} =.070$) and application level ($F(1,75)=4.08$, $p=.047$*, $\eta^2_{\footnotesize\textit{partial}} =.052$). Traces of learning strategies were related to higher coding correctness (organizing: $r=.553$, $p<.001$***; elaborating: $r=.639$ $p<.001$***) and higher learning outcomes (organizing: $r=.400$, $p<.001$***; elaborating : $r=.404$, $p<.001$***). Based on process data, users in the group with both aids had a $5.18$-times higher chance ($p=.020$*) of avoiding certain typical implementation mistakes.

\paragraph{Discussion}
Implementing dynamic linking and highlighting through source location tracking is a promising approach to support novices to develop a better semantic understanding of the domain specific language. Depending on the coding tasks different mapping aids might be effective. Based on traces of learning strategies while programming, adaptive interactive programming environments might be developed to support users individually.

\end{abstract}

\begin{keyword}
\texttt domain specific languages \sep cyber-physical systems \sep novices \sep live programming \sep source location tracking \sep bidirectional linking \sep multiple representations \sep mapping aids \sep multimethod approach \sep traces of learning strategies
\end{keyword}

\end{frontmatter}


\section*{Highlights}
\begin{itemize}
    \item This study investigated the effects of highlighting and dynamic linking between a program editor and a graphical preview in an interactive programming learning environment.
\item We found beneficial effects of dynamic linking on code comprehension and application.
\item Combining both aids compared to highlighting resulted in higher transfer performance.
\item A synergetic effect of both mapping aids compared to dynamic linking was found for programming the first mission.
\item Traces of learning strategies (organizing, elaborating) were related to higher performance.  
\item Mapping aids reduce the odds of typical errors.
\item The best performing novices showed heterogeneity in terms of their characteristics
\end{itemize}
\newpage
\section{Introduction}
\label{sec:introduction}

The programming and integration of cyber-physical systems (CPS) -- such as stationary or mobile robots -- is often performed by domain experts rather than software engineers \citep{rossano.2013}. In rescue operations, for example, drones can help to better assess the extent of an accident, simplify the coordination of an operation, or make an important contribution to the rescue of survivors \citep{roldan2021}. The low-level programming of such a system, i.e., basic control and communication capabilities, is provided by the manufacturer. However, high-level programming of the mission objectives must be done on site by the drone operator -- typically not a programming expert. For this purpose, it is necessary that the programming and learning environment (PLE) for that system is easy to operate and that a basic understanding can be learned quickly. Since many CPS operators do not have any special programming skills, the question arises of how to facilitate the handling and acquiring basic knowledge of these types of systems for novices.

To  overcome initial barriers while interacting with CPS or learning to code, the design of the provided PLE is crucial \citep{Prather.2018}. 

In PLEs different ways to display information are available: depictive and descriptive presentation formats \citep{schnotz2010}. On the one hand descriptive presentation formats such as text or code expressions can be provided. On the other hand depictive formats can be chosen such as pictures or animations (e.g. a preview).

For instance, PLEs for novices might simplify programming by presenting the code in a way that is not only descriptive but contains additional depictive components. For example, graphical programming languages, e.g., based on combining puzzle pieces that each provide a small mission element  \citep[e.g. Blockly;][]{pasternak2017tips, Price.2015}, can be employed instead of solely textual (descriptive) programming languages. Combining intuitive symbols or affordances that imply the effect of the programming code (e.g. puzzle pieces in Blockly) with the descriptive expressions or code components might facilitate programming \citep{Pot.2009}. These approaches seem very promising to lower the initial barrier for novices to code or control CPS. Despite the graphical (depictive) component of e.g. Blockly, this presentation format can still be categorized as textual, as the main information (code element) is presented as text and therefore in a descriptive form.

Based on the literature and learning psychology models, the advantages of depictive presentation formats or their combination with descriptive formats are often recommended for the design of (learning) environments \citep{Schnotz.2003,Ainsworth.2014,Mayer.2014}. For this reason, PLE oftentimes incorporate multiple representations such as a code editor with other views, e.g. a simulation, a graphical preview of the result or other live programming features \citep{DBLP:conf/vl/BurnettAW98,Tanimoto.2013,rossano.2013,BT16b}.

Including multiple representations in PLEs by combining descriptive components such as a program editor (e.g. Blockly) and a depictive representation (e.g. a 3D preview) might challenge the novices to map information between the different formats \citep{Ainsworth.1999, Seufert.2019}. For instance, they need to understand the individual elements but also the relationship between the code and the preview. Only when these relationships are understood, either by linking related elements or by understanding corresponding processes, can an advantage be gained from the use of multiple representations \citep{Patwardhan.2017, Rey.2011, Mayer.2018}. To ease the mapping of multiple representations, corresponding elements can be \textit{highlighted}, or \textit{dynamic linking} between the different representations can facilitate the understanding of the corresponding processes \citep{fries2021, Seufert.2019, Hempel., vanderMeij.2006}. The beneficial effects of mapping aids have been examined in different contexts for decades \citep[e.g.][]{Gentner.1983,Brunken.2005,fries2021}.
A transfer of these principles, which have long been well researched, seems promising in the context of basic CPS knowledge acquisition.

In the CPS domain, novices can be directly supported in the handling of certain programming tasks in an interactive PLE, such as programming a flight trajectory for a quadcopter. In detail, facilitating the interaction with the PLE, understanding of the individual code elements and the underlying processes can be eased by the mentioned mapping aids (highlighting \& dynamic linking). If, during the interaction, the semantics of the language become more transparent -- i.e. the user understands, what the individual elements do and how they interact -- a long-term understanding of the concepts can emerge and novices might learn to apply the concepts when solving future problems \citep{Hadwin.2007, Jeske.2014}. Accordingly, the mapping aids should result in higher \textit{program correctness}. Furthermore, such mapping aids could also influence the understanding of the code components and thus the resulting \textit{learning outcome} \citep{Patwardhan.2017, sobral2021b}. 

In detail, our study aimed at answering the following research questions (RQs):

\begin{itemize}
    \item [\emph{RQ1}] Do mapping aids increase program correctness?
    \item [\emph{RQ2}] Do mapping aids have a positive effect on learning outcomes?
    \item [\emph{RQ3}] How are learning strategy traces and program correctness related?
    \item [\emph{RQ4}] How are learning strategy traces and learning success related?
    \item [\emph{RQ5}] Do those mapping aids reduce typical errors?
    \item [\emph{RQ6}] What are the characteristics of successful novices?

\end{itemize}

We conducted a 2x2 between-subject design study with 82 participants to gain
insights into these different mapping aids (highlighting and dynamic linking) and their impact on performance under consideration of participants' characteristics. We focus on the robotics domain and especially aim to support programming novices, i.e., users with no or only basic programming knowledge, with our PLE and our proposed mapping aids. We therefore selectively recruited participants matching these criteria and refer to these participants as novices throughout the paper.

By implementing a multi-method approach, we investigated the effects of mapping aids on the program correctness of three different programming tasks (quadcopter missions; RQ1) as well as on learning outcomes (knowledge, comprehension, and application level; RQ2). Based on process data, we examined the effects of mapping aids on traces of learning strategies (RQ3 \& RQ4), typical errors during programming (RQ5) and characteristics of successful novices (RQ6). 

Based on our findings for RQ1 and RQ2, support by highlighting enabled novices to achieve better program correctness (mission 1: $p=.008$, $\eta^2_{\footnotesize\textit{partial}}=.091$) while support by dynamic linking increased learning outcome (comprehension: $F(1,75)=5.61$, $p=.020$*, $\eta^2_{\footnotesize\textit{partial}} =.070$ \& application: $F(1,75)=4.08$, $p=.047$*, $\eta^2_{\footnotesize\textit{partial}} =.052$). Overall, the best performance was found when both mapping aids were available. 
For RQ3 and RQ4, traces of learning strategies (organizing, elaborating) were positively related to both, program correctness (organizing: $r = .553$, $p < .001$***, elaborating: $r = .639$, $p < .001$***) and learning outcome (organizing: $r = .400$, $p < .001$***, elaborating: $r = .404$, $p < .001$***). Participants in the group with both aids had a $5.18$-times higher chance ($p=.020$*) of avoiding certain typical implementation mistakes (RQ5). Successful novices (based on the program correctness) had heterogeneous characteristics, i.e. we did not identify a common characteristic (or combination) leading to programming success (RQ6).

The remainder of this paper is structured as follows: Sections~\ref{sec:background} and \ref{sec:relatedwork} summarize the technical and psychological background and give an overview of the state of the art and related work. The hypotheses derived from the related literature and research questions as well as a brief overview of the study is given in Section~\ref{sec:presentstudy}. In Section~\ref{sec:method}, we describe the study design, methods and procedures in greater depth. The results of our study are then presented in Section~\ref{sec:results} before discussing the implications of these results, as well as threats to validity in Section~\ref{sec:discussion}. In Section~\ref{sec:conclusion}, we conclude and summarize our results and give an outlook into possible future work.

\section{Background}
\label{sec:background}
First, we will outline the implementation of the PLE before considering underlying cognitive processes for the different components of the PLE.

\subsection{Live programming}
Live programming tools or environments give direct and continuous feedback to the user by analyzing and visualizing the effects of changes to the program while editing it \citep{Tanimoto.2013}. 
Established and ubiquitous concepts like syntax highlighting -- i.e. analyzing the source code and highlighting or different coloring of keywords, identifier, literals, etc. -- are not considered live. In contrast, more advanced live programming concepts visualize execution behavior, e.g., the value of variables during execution (live evaluation) or give a preview of the textual or graphical output of the program that changes with each editing operation.

Especially graphical output or output that has a spatial or physical dimension, e.g., visualizations of the environment, is well suited for liveness due to the natural and intuitive visualization that fosters program understanding \citep{Tanimoto.2013}. Programming CPS, such as robots, can profit from live programming as the feedback loops when testing new code are especially long: transferring and running programs on the robot and observing the results is often time consuming and might even pose safety risks to the operator \citep{Campusano.2017}. In contrast, a live preview provides immediate feedback and might even incorporate data from the physical environment or a simulation to help identify potential errors or safety hazards.

The direct feedback and support in visualizing the program execution is also an important aspect of programming environments focused on teaching programming fundamentals to kids or novices. In addition, live programming features help forming mental models of the program and language semantics and can lead to more experimentation \citep{Kang.2017}.

\subsection{Source location tracking}
\label{sec:slt}
Live programming features can be augmented through \emph{bidirectional editing} features \citep{Hempel.}, allowing program manipulation both through code and the program result.
\emph{Source location tracking} (SLT) \citep{BT16b, Witte.2019} is an implementation pattern that links program output to its source by tracking data flow during execution. This allows implementing dynamic links between code and its execution result without providing additional domain knowledge.
Thoughout the paper, SLT is used refer to the implementation technique and bidirectional editing synonymously while dynamic linking and highlighting refer to concrete PLE features using SLT.

\begin{figure}[htbp]
    \centering
    \includegraphics[width=0.7\textwidth]{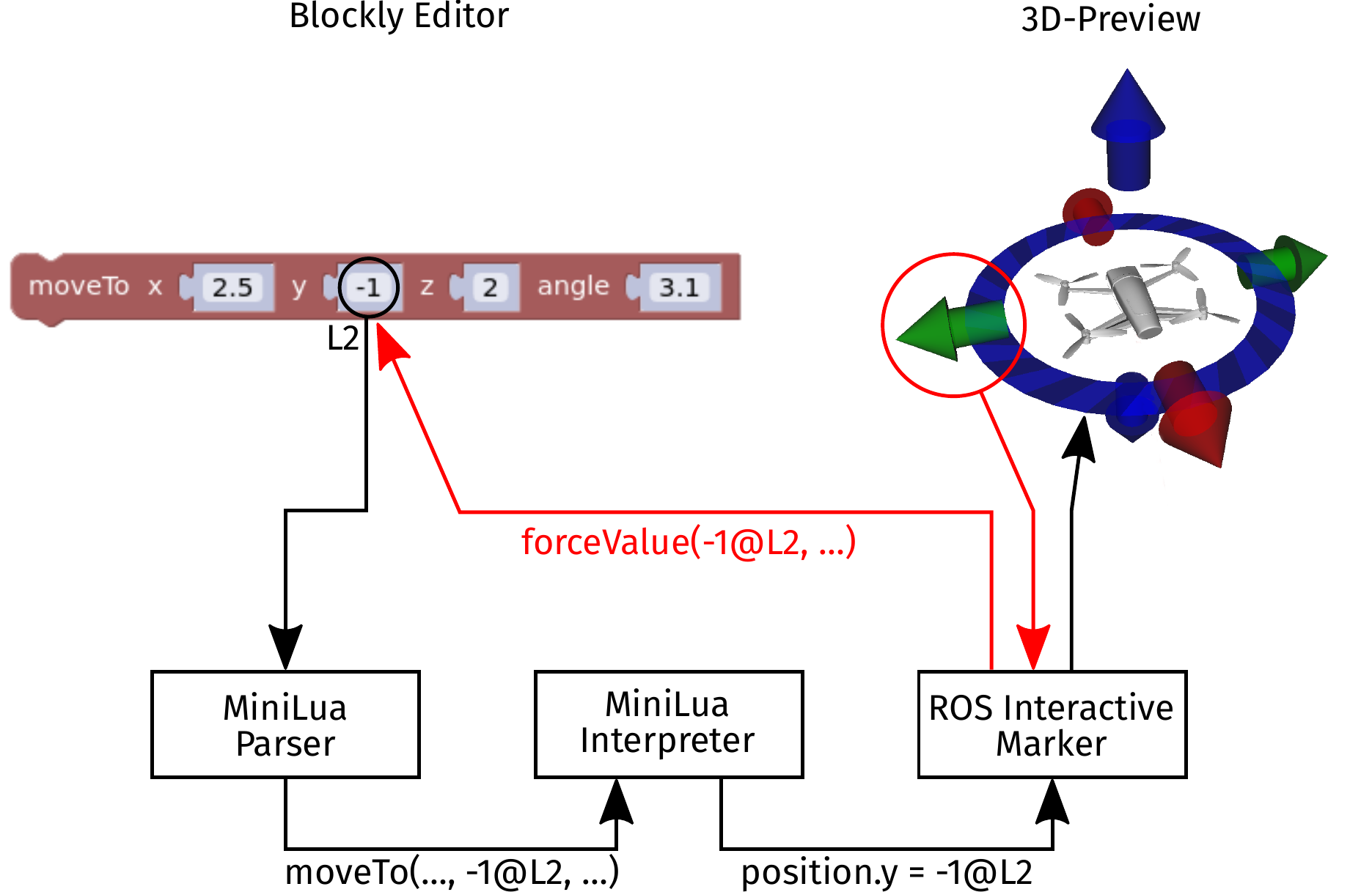}
    \caption{Interactive waypoint using SLT location information to change the corresponding value in the Blockly editor.}
    \label{fig:slt}
\end{figure}

Imagining a quadcopter instructed to fly to a certain waypoint using a \emph{moveTo} block as shown in Figure~\ref{fig:slt}, the code location/block responsible for the quadcopter's position along the $y$-Axis (L2) is known throughout the complete execution and rendering process of the preview.

This way, SLT complements live programming by automatically forming links between code and visualization as a side effect during the generation of the visualization. As the relation between visualization elements (the preview of the quadcopter on the right) and the code (the block on the left) is known, code changes corresponding to user changes in the visualization can be proposed or directly applied. Changing the position of the preview using the green arrow along the $y$-Axis uses the saved location information associated with the value and tracks it back to its origin (shown in red). The code change is then directly reflected in the live visualization giving immediate feedback to the user \citep{Hempel., Witte.2019}. For example, if the waypoints define a square, the user might move one of the waypoints to enlarge the square and the constants that define the size of the square are automatically adapted in the code.

As this linking between code and preview is realized on the fundamental level of the programming language, no semantic knowledge of the \emph{moveTo} block or the waypoints in the preview is needed. Any (intermediate) result of a computation in the user's program can be automatically tracked back to the literal values used to compute it by analyzing and following its recorded history. This process is mostly application and domain independent and was previously applied in e.g. drawing programs \citep{Hempel.}, live evaluation in text editors \citep{BT16b, hempel2020tiny}, HTML and Markdown editors \citep{Mayer.2018}.

Due to tracking data flow during execution and zero domain knowledge, SLT can be easily integrated into existing live programming environments but introduces some runtime overhead. Compared to static analysis approaches, SLT can produce inaccurate results in the presence of aliasing or changing dataflow.

\begin{figure}[htbp]
    \centering
    \includegraphics[width=.7\textwidth]{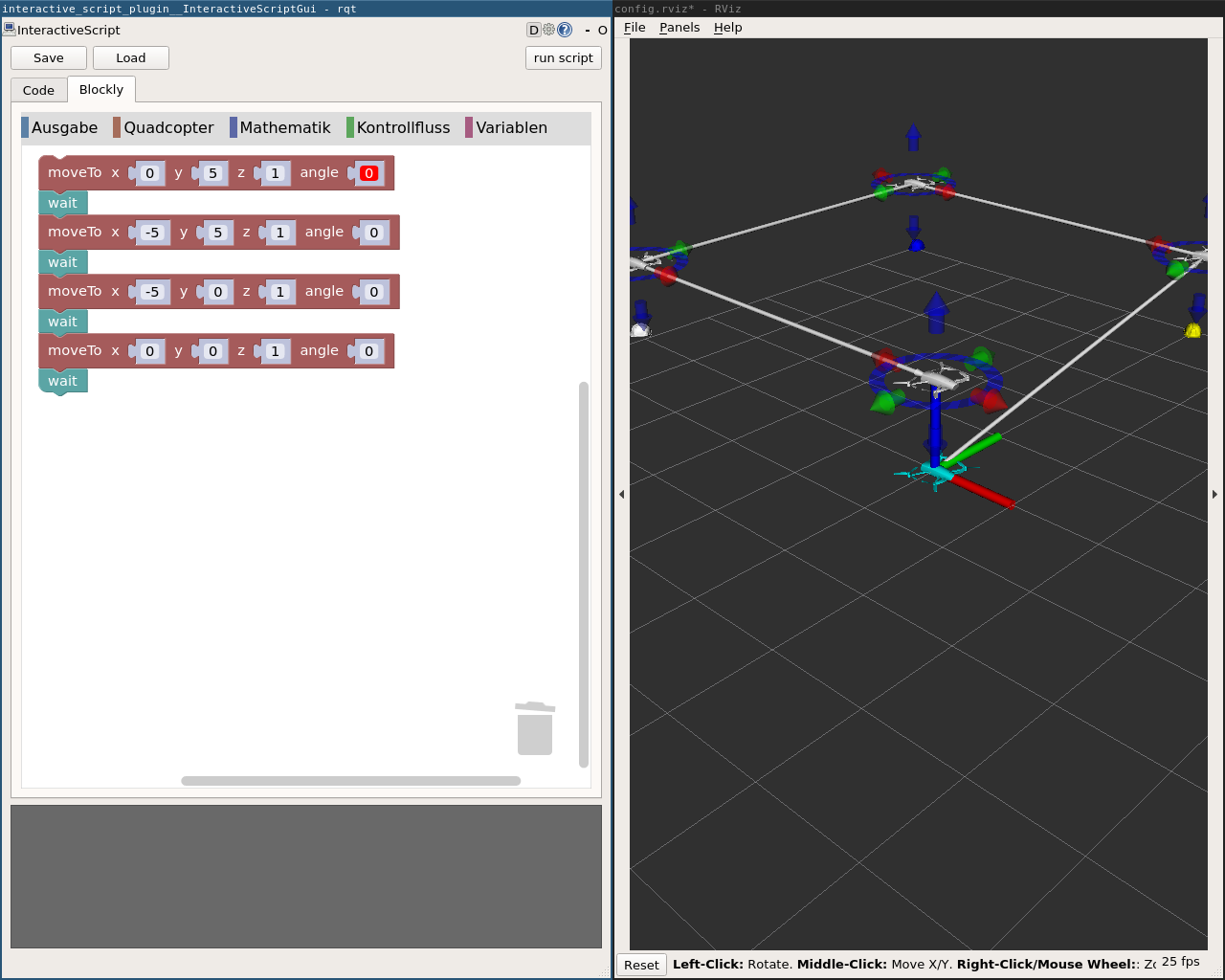}
    \caption{interactive\_script editor with visual program editor and interactive preview.}
    \label{fig:interactive_script}
\end{figure}

The \emph{interactive\_script} environment \citep{Witte.2019} incorporates live programming and visual programming features to easily program simple quadcopter missions (cf. Figure~\ref{fig:interactive_script}) and is adapted to serve as a PLE for this study. A Lua-based domain specific language, designed around a strictly imperative, high-level programming paradigm is used to program quadcopter missions using only a few basic functions and a simple program flow without callbacks or events. To further help novice programmers, a \emph{Blockly}-based visual program editor reduces the syntactic complexity of the language and encourages experimentation \citep{kolling.2017}. Each basic function is represented by a puzzle block -- connecting it to other pieces creates a syntactically correct program. A live visualization of the currently edited program is shown in a 3D preview in real time, displaying waypoints and the flight path of the quadcopter. The interactive preview shows all waypoints and flight paths by ignoring the time dimension. However, users can start a simulation of the quadcopter that shows an animated 3D model of the quadcopter executing the mission in real time.

\subsection{Underlying cognitive processes in the context of programming and learning environments}
After providing insights into the technical details of the PLE, the underlying cognitive learning processes when interacting and learning in this environment, will be described in more detail in the following. The present PLE includes multiple representations such as programming elements (descriptive) and animated previews (depictive). According to well-known theories such as the \textit{Cognitive Theory of Multimedia Learning} \citep{Mayer.2014, Mayer.2020} and the \textit{Integrated Model of Text and Picture Comprehension} \citep{Schnotz.2003}, humans first process descriptive representations such as the programming code and depictive representations (e.g. a 3D-preview) via different channels. A more in-depth semantic processing takes place after processing the content superficially. If enough cognitive resources are available, processing and learning is successful and new information is integrated and linked in a complex, analog \textit{mental model} \citep{Schnotz.2003}.  However, these theories do not refer to the challenge of mapping and integrating the processed information in more detail. This mapping is particularly important when learning with multiple representations. Learning with multiple representations can be challenging because working memory might be (over-)loaded by the presentation of many different pieces of information \citep{van1998learning}. In particular, learners with little prior knowledge cannot manage working memory capacity by chunking and are particularly affected by this problem \citep{kozma1997}. These challenges might result in novices using simulations in a haphazard and exploratory rather than systematic way \citep{Patwardhan.2017,de_jong_1998}.

The specific challenges of learning with animations, in this case the activation of the quadcopter flight simulation in the 3D-preview, are made explicit in the \textit{Animation Processing Model} \citep{lowe2008learning}. Here \cite{lowe2008learning} describe that especially novices in animation learning tend to focus on localized perceptual exploration by parsing individual event units and are not able to develop more complex mental models with a more global understanding. This means that novices are not able to form dynamic micro chunks and to understand more precisely the spatiotemporal structures in the specific context \citep{lowe2008learning, lowe2014animation}. In other words, the presentation of animations or multiple representations does not in itself offer a general advantage, but that the elements in the PLE and their combination and linkage should be carefully chosen.
 
\subsection{Mapping multiple representations}
When multiple representations are included in a PLE, they can have different roles \citep{Ainsworth.2014}: a) complementing each other b) limiting each other's interpretation, or c) promoting deeper understanding. In the case of interactive PLEs, i.e. PLEs that the learner can directly interact with or manipulate, multiple representations often take the third role, with the goal of fostering a deeper understanding \citep{Rey.2011}. To facilitate this, additional aids are provided in such PLEs to help locate corresponding information and thus simplify mapping \citep[e.g.][]{fries2021, vanderMeij.2006}. For instance, findings that also make use of psychophysiological methods (eye-tracking) indicate that especially when using animated simulations such as a 3D-preview, the use of certain visual cues (dynamic signals) can be helpful to process relevant information and to develop a more global understanding \citep{boucheix2013cueing}. These helpful cues can be differentiated based on the processes they are related to: Different mapping processes for depictive and descriptive presentation formats have been described by \cite{Gentner.1983}. Mapping processes can be distinguished into element-to-element and relation-to-relation mapping processes \citep{Gentner.1983, Seufert.2019}. The term element-to-element mapping refers to the element-based connection reflecting that novices are able to connect different components on a rather superficial level \citep{fries2021, Gentner.1983}. In detail, highlighting of dependent values as a form of element-to-element mapping enabled our participants to click on a waypoint to highlight all literal values that may influence the position. In contrast, the term relation-to-relation mapping refers to the finding of similarities on the semantic level and thus requires the novices to have a deeper understanding of the elements and processes that need to be mapped and integrated \citep{Patwardhan.2017, Gentner.1983, Seufert.2019}.
In our PLE, dynamic linking as a form of relation-to-relation mapping, enabled participants to directly move waypoints through the mission preview. 

\subsection{Novices' characteristics}
Performance in solving a task and understanding its components depends not only on the design of the PLE, but also on certain characteristics of the novices. 

Based on previous findings, relevant individual characteristics were considered as covariates (see Section~\ref{IT}). For instance, the importance of prior knowledge for successfully conducting specific programming tasks was outlined as well as the importance of this factor on learning outcome \citep[e.g.][]{richter2019}. To plan the flight trajectory of the drone, figural intelligence or reasoning might impact the final success \citep{Helmlinger.2020}. Furthermore, when novices interact with complex tasks, their level of need for cognition might be crucial for the question of whether they succeed and develop a deeper understanding of the learning content \citep{Coutinho.2005, colling2022}.

\subsection{Performance assessment}
For comparing different PLEs and their effects on critical outcome variables, several promising criteria can be considered based on theory and previous findings. These criteria and their embedding in theory, taking into account empirical findings, are described in more detail in this section.
Different cognitive sub-steps while interacting and learning in a PLE can be distinguished and might only become apparent when using a differentiated measurement of learning outcome or task performance.
The performance of novices handling programming tasks (program correctness) can be used as a criteria for evaluating how novices can be successfully supported by mapping aids (RQ1). In order to solve the tasks successfully and achieve good performance in program correctness, novices must develop a basic understanding of the individual components and processes. This understanding is already implicitly measured by assessing the program correctness. 

In addition, the knowledge of e.g. definitions or functionalities of the individual code elements can be explicitly assessed after the interaction in the PLE. Based on the impressions and insights gained during the processing of the quadcopter missions, novices could also develop an understanding of the concepts, which they can later apply to new problems. In other words, they not only successfully master the missions, but also learn something about the components and processes. Based on \cite{Bloom.1956} the three levels of learning outcome knowledge, comprehension, and application can be distinguished (RQ2). The knowledge level can be achieved when simply storing a superficial representation of the learning content. In contrast, users need to semantically process the information to succeed on the comprehension and the application level \citep{sobral2021b}. More specially, the ability to contrast different concepts or code components and understand the relationship between two code elements represents the learning outcome at the comprehension level. When users have integrated the new information into the more global and analog mental model, they are able to deduce more complex consequences and interactions between the different components as well as apply the concepts to different problems \citep{mayer2002.fostering}. 

Furthermore, cognitive learning strategies such as organizing and elaboration can be derived based on the learning behavior by using process data \citep{HADWIN2021, Jeske.2014}. In addition, the learning behavior can contain traces of metacognitive strategies, which can map, for example, planning or monitoring during the programming process \citep{HADWIN2021,Jeske.2014,WildKPSchiefeleU.1994}. Hence, traces of certain strategies while working on the programming tasks could already provide information about whether users use adequate strategies when solving certain tasks and achieve a higher level of performance with these strategies (RQ3 \& RQ4). 

Furthermore, typical errors can be identified and by this more insights into the effects of the mapping aids might result (RQ5).

In addition to the question of which aids in PLEs have a beneficial effect, it is also important to differentiate the extent to which only certain novices benefit from the aids offered or whether a certain profile of experiences and characteristics already promises greater success \citep{snow1989}. On the one hand, this can be done by including the critical characteristics in the quantitative analyses. On the other hand, these profiles can also be examined again in terms of content, for example by taking a closer look at the profiles of the most successful novices (RQ6).

\section{Related work}
\label{sec:relatedwork}

Due to the fact, that mapping aids in PLEs are an interdisciplinary research field, very different keywords were used and consequently, finding related studies and empirical results is not self-evident. 
In this section, we will provide insights into related work from both, software engineering and learning and instruction research. While publications in the field of software engineering oftentimes focus on technical aspects, the learning and instruction perspective focuses on empirical evaluation and attributing results to related cognitive processes. 

\subsection{Live programming features in programming environments for robotics}\label{rwIT}
Particularly, live/online PLEs offer different options to implement supportive elements such as mapping aids. Multiple representations or live programming features are commonly used in programming environments for robotics that are targeted at programming novices. Choregraphe \citep{Pot.2009}, the programming environment for the NAO humanoid robot uses a data flow oriented graphical programming model paired with a poseable preview of the robot as well as a choreography timeline.

Sketch 'n Sketch \citep{Hempel.} uses similar techniques to source location tracking to create an interactive visual preview with multiple application cases like HTML, SVG and markdown editors using functional domain specific languages as a linked, textual representation.

Quantitative, empirical studies of usability and learning aspects of these environments, however, are rarely done. \cite{berenz2014} compare their own, declarative behavior specification language with the flow based programming approach of Choregraphe in a user study with 17 participants from a professional or educational background. The study focuses on their alternative programming paradigm \emph{Targets-Drives-Means (TDM)} and uses Choregraphe only as a state-of-the-art baseline and does not consider the underlying processes related to multiple representations or mapping aids. 

The bidirectional programming/SLT features of Sketch 'n Sketch, i.e. the ability to manipulate code directly through a preview of its result (cf.~Section~\ref{sec:slt}) -- to our knowledge -- were never evaluated in a user study. However, the usability of the interface of the D\textsc{euce} program editor, implemented in Sketch 'n Sketch, was evaluated through a small user study with 21 participants \citep{Hempel.2018}. The proposed structural selection mechanism is compared to a baseline version of the same tool offering only conventional text selection features. As this study does not cover code manipulation through the interactive visualization, it provides no results on the effects of mapping and linking of multiple representations. 

Other case studies focusing on specific aspects, e.g., block-based robot programming in a learning context for children \citep{Sutherland.2018} underlining the importance of instant, easy to understand and robust feedback. \cite{Winterer.2020} use Blockly in an industrial robot programming context showing that even complex programs can be expressed in Blockly, and that the Blockly interface provides similar or better understandability or maintainability than traditional flowchart-based visual languages. \cite{Price.2015} compare block based interfaces to textual interfaces in a novice programming environment. While the perceived difficulty does not change, participants using the block based interface spent less time off task and completed the exercises faster. While the used interfaces often include some form of preview, they do not consider features linking both views bidirectionally or measure the performance impact of the linking of both views. \cite{Campusano.2019} evaluated live programming features to program robot behavior through state machines in an experiment. Contradicting their assumptions, the live environment did not outperform the non-live, state-of-the-art baseline but participants preferred the live environment. 

\subsection{Evaluations of mapping multiple representations in interactive environments}
 Many different ways to facilitate mapping in interactive PLEs have been reported and evaluated in terms of their effects on cognitive processes and learning success. Some older studies, for example, used graphical techniques to mark corresponding elements \citep[inter-representational hyperlinks:][]{Brunken.2005, Seufert.2007}. In this case, relevant elements within a representation were marked as hyperlinks and users had the possibility to click on these inter-representational hyperlinks to get hints about corresponding elements in other representations. A beneficial effect of this implementation of element-to-element mapping aid was described when comparing results of the experimental group with a control group \citep{Seufert.2007}. 
 
More recently, the approach of linking multiple representations, allowing manipulation of a virtual simulation, was studied in more detail. For instance,  \cite{Rey.2011} compared different mapping aids in a study (text fields, scroll bars, or drag-and-drop) as possibilities that allowed the modification of the parameters of a virtual simulation. These three interactive elements were used to dynamically link the different (multiple) representations with the simulation. Hence, they represent an implementation of relation-to-relation mapping aids, as they do not simply outline related elements but make related processes more transparent \citep{Gentner.1983}. Here, positive effects of some interactive elements (scrollbars and drag-and-drop) were found on the transfer level, but not on the knowledge level compared to the text field condition. As already outlined in the section before (see Section~\ref{rwIT}) by the example of Sketch 'n Sketch, this approach of dynamic linking between a virtual simulation and a textual representation has been further extended \citep{Mayer.2018, Hempel.}.

Bidirectional linking of multiple representations -- similar to the dynamic linking feature in our PLE -- 
has been implemented in some studies as one possibility to implement relation-to-relation mapping aids. In this case, users were able to manipulate any of the related components to cause a change in the respective other representation \citep{Hempel.}. For instance, \cite{Patwardhan.2017} uncovered the beneficial effects of bidirectional dynamic links as mapping aids, not only for interacting with a PLE, but also on higher levels of learning outcome (transfer).

These (heterogeneous) findings illustrate that live programming features per se are not an advantage and that it is necessary to obtain more precise information in larger empirical studies on which type of mapping aids are particularly helpful, which underlying processes are influenced by these mappings aids and how, and which boundary conditions and characteristics of novices play a role in this.

\section{Research Hypotheses}\label{sec:presentstudy}

\begin{figure}[htbp]
    \centering
    \includegraphics[width=0.9\linewidth]{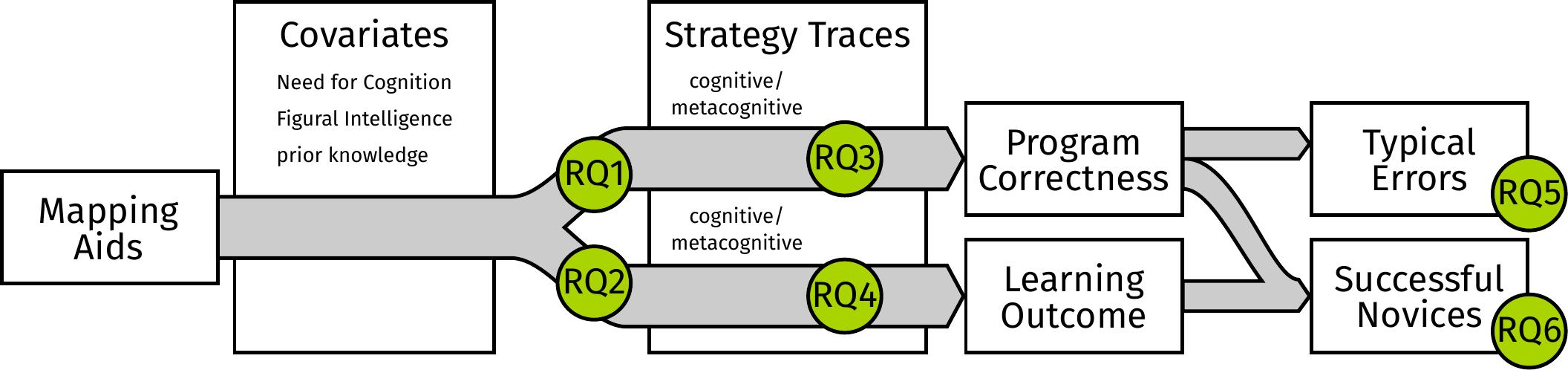}
    \caption{Relevant factors and related research questions.}
    \label{fig:concepts}
\end{figure}

Given that this study and its methodology are not only based on the problem statement from the perspective of software engineering, but also on findings from the research field of learning and instruction, as well as important cognitive psychology findings, the specific hypotheses are theory-based and were postulated under consideration of previous empirical findings. As recommended by previous publications, we used a multi-method approach to assess performance in our study, that offered further insights into the different effects of the implemented mapping aids \citep[e.g.][]{Makransky.2019}. 
In the following, we present based on our research questions (RQs; Figure~\ref{fig:concepts}), our more specific hypotheses. Our RQ1 and RQ2 look into the effect of the mapping aids to \emph{program correctness} and \emph{learning outcome}; traces of cognitive and metacognitive strategies in this process were considered in RQ3 and RQ4. Additionally, RQ5 investigates typical error in the different experimental groups. Finally, RQ6 provides more insights into the profiles of successful novices. 

\subsubsection*{RQ 1: Do mapping aids increase program correctness?}
Our first research question explored the effect of mapping aids (highlighting \& dynamic linking) on program correctness. By highlighting, users' visual attention was guided and they could see corresponding elements. Hence, an unnecessary visual search was reduced and selection processes were eased as the respective elements needed no further consideration whether they are corresponding or not \citep{Ozcelik.2010, fries2021}. 

\begin{itemize}
    \item[\emph{H1a}] We hypothesized a beneficial effect of highlighting on program correctness of the quadcopter missions.
\end{itemize}

\noindent
Relation-to-relation mapping aids supported the users as they outlined the underlying processes by showing one possible solution for the change in the preview by altering the respective coordinates in the code \citep{Seufert.2019, Mayer.2018}. 

\begin{itemize}
    \item[\emph{H1b}] We hypothesized a beneficial effect of dynamic linking on program correctness of the three quadcopter missions.
\end{itemize}

\noindent
As both highlighting and dynamic linking might ease the programming process, using both mapping aids in combination was expected to show a synergetic effect \citep{Rey.2011}:

\begin{itemize}
\item[\emph{H1c}] We expected a synergetic effect of both mapping aids compared to a single mapping aid on the program correctness of the different missions.
\end{itemize}

\subsubsection*{RQ 2: Do mapping aids have a positive effect on learning outcomes?}
Some previous studies showed beneficial effects of mapping aids compared to control groups without additional help to connect the different representations in the PLE \citep{Rey.2011, Patwardhan.2017}. Based on theoretical considerations, the chosen mapping aids trigger specific cognitive learning processes. Element-to-element mapping (highlighting) was assumed to ease finding corresponding elements and by this, the integration of the different representations \citep{fries2021}.

\begin{itemize}
    \item[\emph{H2a}] We hypothesized the strongest beneficial effect of highlighting on the comprehension and application level of learning outcome compared to the knowledge level of learning outcome. 
\end{itemize}

\noindent
Furthermore, our relation-to-relation mapping aid (dynamic linking) outlined the connectedness of different components in the PLE and the consequences of manipulating one component (the live preview) onto another component (the code editor). This effect was expected to reflect mostly the learning outcome on comprehension level \citep{Patwardhan.2017}.

\begin{itemize}
    \item[\emph{H2b}] We hypothesized the strongest beneficial effect of dynamic linking on the comprehension and application level of learning outcome compared to the knowledge level of learning outcome. 
\end{itemize}

\noindent
Based on the presented theories and the previous research, it was plausible, that users would benefit more from the combination of both mapping aids than from one mapping aid on the different levels of learning outcome \citep{Patwardhan.2017, Rey.2011}.

\begin{itemize}
    \item[\emph{H2c}] We expected a synergetic effect of both mapping aids compared to a single mapping aid on the different levels of learning outcome. 
\end{itemize}

\subsubsection*{RQ 3: How are learning strategy traces and program correctness related?}
In our third research question, we considered the relationship between behavioral patterns in the PLE and program correctness. As mentioned above, traces of cognitive and metacognitive learning strategies such as organizing and elaborating can be inferred from process data \citep{HADWIN2021, Jeske.2014, WildKPSchiefeleU.1994}. 
Based on the literature, it is expected that purposeful interaction in the PLE using strategies will lead to better performance in terms of program correctness \citep{de_jong_1998,Patwardhan.2017}.

We hypothesized a positive correlation of program correctness and …
\begin{itemize}
\item[\emph{H3a}] cognitive organizing strategies traces. 
\item[\emph{H3b}] cognitive elaboration strategies traces. 
\item[\emph{H3c}] metacognitive planning strategies traces. 
\item[\emph{H3d}] metacognitive monitoring strategies traces. 
\end{itemize}

\subsubsection*{RQ 4: How are learning strategy traces and learning success related?}
According to the third research question in which a positive effect of strategy use on program correctness was expected, a positive effect of these strategy traces on overall learning success was also expected \citep{HADWIN2021, WildKPSchiefeleU.1994, Patwardhan.2017}: 

We hypothesized a positive correlation of overall learning outcome and …
\begin{itemize}
\item[\emph{H4a}] cognitive organizing strategies traces. 
\item[\emph{H4b}] cognitive elaboration strategies traces. 
\item[\emph{H4c}] metacognitive planning strategies traces. 
\item[\emph{H4d}] metacognitive monitoring strategies traces. 
\end{itemize}

In the following research question, we explore the process data to gain further insights into typical errors and novices characteristics that might impact performance.

\subsubsection*{RQ 5: Do those mapping aids reduce typical errors?}
In our next research question, we aimed at gaining deeper insights into what were specific challenges while programming the missions and how typical errors were distributed in the different experimental groups with respective mapping aids. This was in line with prior publications analyzing difficulties of novices while programming \citep{Prather.2018, Patwardhan.2017}.

\begin{itemize}
\item[\emph{H5}] Based on these previous findings, we expected the odds of making typical programming errors in the first mission to differ between groups. 
\end{itemize}

\subsubsection*{RQ 6: What are the characteristics of successful novices?}
In addition to RQ5, which focused on traces of learning strategy-related behavior patterns, the sixth research question focuses on higher-level behavior patterns and characteristics of the ten most successful novices. Based on a content based analysis of this subgroup, we will identify particular characteristics that promise successful solution strategies. 
\begin{itemize}
\item[\emph{H6}] We expected that the most successful novices would differ in their aptitudes from the overall mean.
\end{itemize}

\section{Method}
\label{sec:method}

In the following section, we provide more details about the procedure (cf.~Figure~\ref{fig:studyoverview}), participants, study design, the PLE including the programming task and the used questionnaires as well as some insights into the data preparation.

\subsection{Procedure}
\begin{figure}[htbp]
    \centering
    \includegraphics[width=0.8\textwidth]{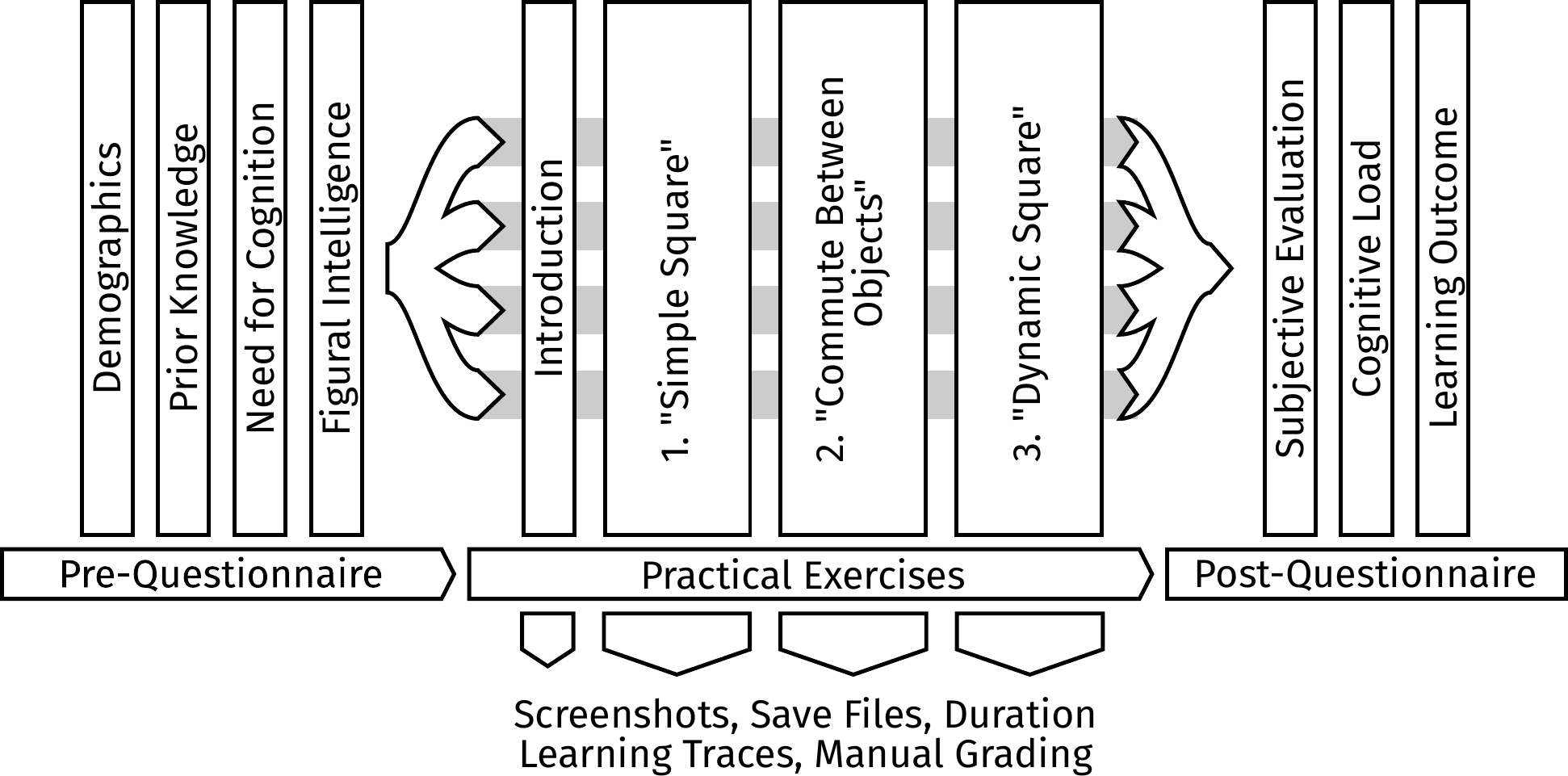}
    \caption{Elements of the study: questionnaires and programming exercises.}
    \label{fig:studyoverview}
\end{figure}

At the beginning of the online study, an overview of the study as well as informed consent was presented to the participant in the online pre-questionnaire (see Figure~\ref{fig:studyoverview}). Participants were aware that they could withdraw the study and the related data at any point in the study without having any disadvantage. The pre-online questionnaire started with demographic questions and the pre-test included the prior knowledge questions. At the end of the pre-questionnaire, a link was included to log into the interactive PLE to code the three quadcopter missions. After finishing the three tasks, users filled out the post-questionnaire. Next, users answered the post-test learning outcome questionnaire and had the opportunity to give feedback on the provided mapping aids if these were available. At the end of the study, they were able to comment on the study and report technical problems during the study. In a separate online survey, they were able to leave their personal information to receive their compensation (a voucher). Overall, the study took about 1.5 hours.

\subsection{Participants and study design}
The sample needed, to investigate the effects of the chosen mapping aids was estimated based on an a priori power analysis by using G*Power version 3.1.9.4 \citep{Faul.2009}. The potential effect size of the chosen mapping aids was calculated regarding the reported results from \cite{vanderMeij.2006} with $f^2(V) = .10$ (medium effect size based on Cohen, 2013); $\alpha = .05$; power $(1-\beta) = 0.9$. By this, we determined the minimum of required participants with $N =91$. Initially, $93$ participants, who were mainly university students in psychology, took part in our interactive online study. Because of technical problems, we excluded eleven participants from further analysis. The remaining $N = 82$ participants ($26.50 \%$ male) were aged between $19$ and $39$ years. ($M_{age} = 25.02$; $SD_{age} =  4.04$). We applied a 2x2 between-subject design by randomly assigning the participants to one of the four different design options: with highlights and dynamic linking ($n = 20$), with dynamic linking ($n = 21$), with highlights ($n = 22$) and the control group ($n=19$). As a dependent variable, the program correctness of three quadcopter mission tasks was assessed. Additionally, learning outcome was measured on three different levels: knowledge, comprehension, and application. While participants were working on the programming tasks, screenshots of the PLE were automatically taken every five seconds to track the progress and gain insight into the problem solving process. The program code and layout was automatically saved with a timestamp after each change. Additionally, participants could create manual save files and were prompted to do so after every exercise. This allowed further insights into strategies and typical errors. The study was part of a bigger research project and this paper focuses on the presented research questions. Figure~\ref{fig:studyoverview} gives an overview of the sequence of the study, the contents of the pre- and post-questionnaires and the measured variables.

\subsection{PLE architecture \& implementation of mapping aids}

We developed an interactive online PLE by adapting the \emph{interactive script} editor described in \citep{Witte.2019} and previously introduced in Section~\ref{sec:slt}. To fit the requirements of the study, the existing editor was modified to 1) include instructions to guide the participants through the different exercises (cf. \ref{sec:instruction_pdf}), 2) be accessible online and without installation through a web browser, and 3) implement the different experimental groups and collect process data to document the implementation process.

\begin{figure}[htbp]
    \centering
    \includegraphics[width=\textwidth]{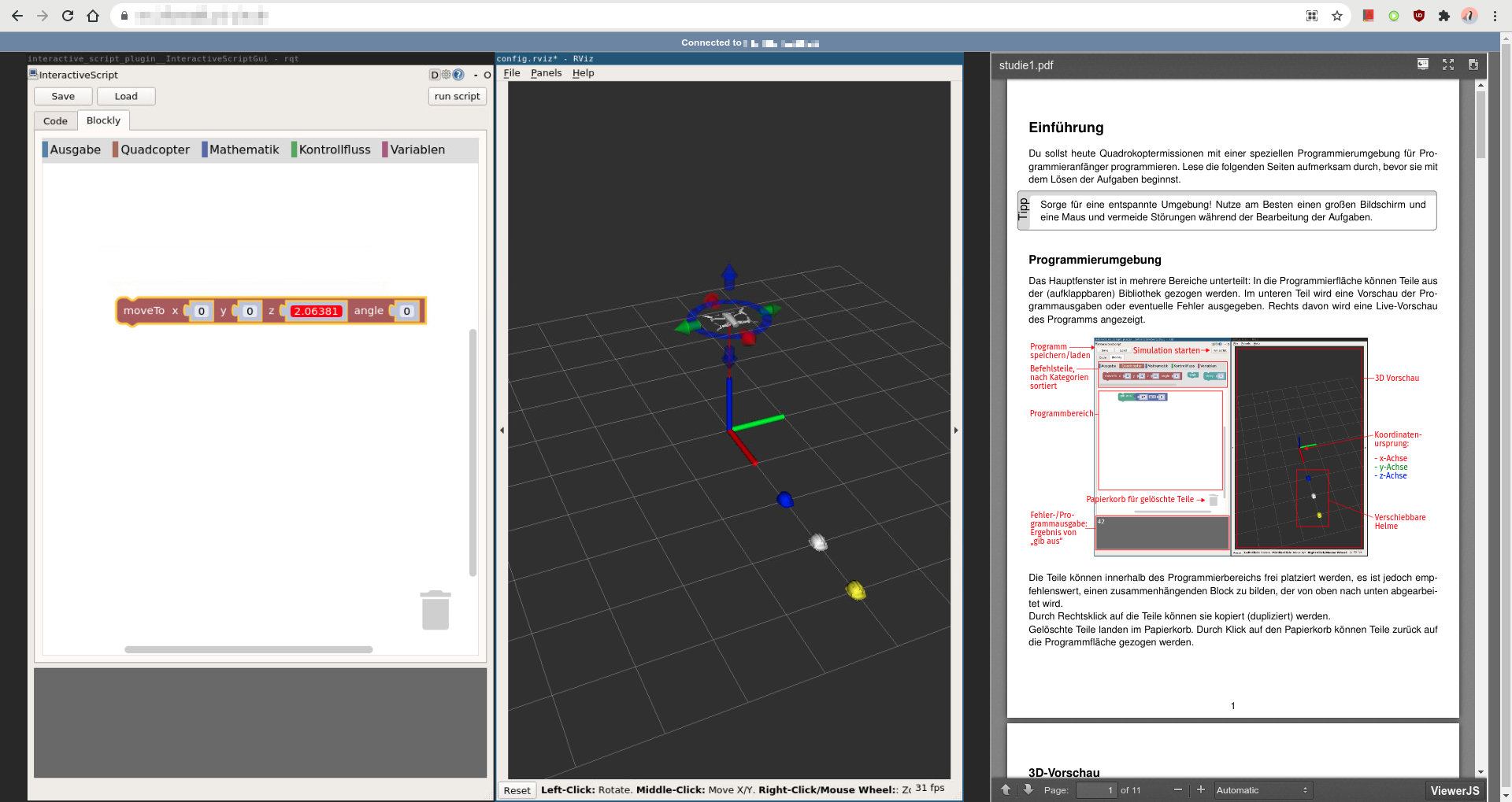}
    \caption{Online PLE as seen by a participant of the study: code editor (left), 3D preview (center), and instruction PDF (right).}
    \label{fig:online_ple}
\end{figure}

The PLE includes different views to display the code for the virtual quadcopter using a block based editor, and a 3D preview of the planned flight trajectory. Additionally, instructions for the programming exercises are displayed alongside the environment (Figure~\ref{fig:online_ple}).

On the left side of the PLE, the Blockly based visual code editor \citep{pasternak2017tips} for a domain specific language to define quadcopter missions was shown. This presentation format allowed the users to create their code through graphical manipulation as it represented program primitives, e.g. values, statements and expressions, as interlocking blocks. New program blocks could be selected from a library and dragged into the program canvas. Additionally, program output was shown in a live console below, allowing users to print and inspect values during development as well as during simulation execution. In the center, an interactive 3D preview for the planned mission of the quadcopter was displayed. Users were able to choose their perspective for the Cartesian coordinate system. To facilitate the orientation in this preview, a grid on the ground plane, the origin, and colored indicators along the x,y, and z-axis were shown. Three colored objects that could be dragged in the 3D view could be used as reference points and their pose queried in the code. The quadcopter pose in each waypoint in the code was displayed in this preview, with lines connecting subsequent waypoints. By pressing the button run script, a simulated quadcopter could be observed while flying along the planned trajectory. Depending on the experimental condition, different dynamic links between the code and the 3D preview were available. 
These different mapping aids are described below in more detail.

In the control group, the planned flight trajectory was displayed in the preview. Users in the group with dynamic linking as relation-to-relation mapping aid were able to click at the way points of the flight trajectory and move these in the virtual space to different coordinates. This dynamic link option was indicated by arrows surrounding each coordinate. While changing the virtual preview, the implemented coordinates in the Blockly code adapted automatically to one possible code for the displayed flight trajectory and the given solution was highlighted in red. Therefore, users were able to observe one specific solution of the given flight trajectory and one code example was provided. 

Users in the group with highlighting as element-to-element mapping aid were able to click on coordinates included in the planned flight trajectory. When clicking on the way points, the related parts in the Blockly code were highlighted in red. The displayed flight trajectory could not be manipulated in the 3D preview in this condition. In the fourth experimental group users were able to use the support by both mapping aids: highlighting and dynamic linking. Therefore, they were able to inspect the relevant input of the code for one given coordinate and to manipulate the flight trajectory in the preview resulting in a changed Blockly code.

Due to the Covid-19 pandemic, the study had to be conducted remotely on the participants' personal computers. To avoid complicated installation procedures, the PLE was executed on a server with a browser-based remote connection tool displaying the application in the browser window and sending inputs to the server. Inevitably, this introduces the risk of connection losses as well as latency issues causing delayed reaction to input. In our testing and through questionnaire items asking for any technical issues during the study, we found these issues affecting only a few participants. If the connection was lost, the page could be simply reloaded with no progress lost as the application still ran on the server. Furthermore, running the application on the server helps controlling the PLE: the application had a fixed resolution that was scaled to the participant's window size, ensuring that each participant had a similar viewport and performance. Screenshots of the PLE were automatically made without the risk to capture private data on the participants personal computer or requiring the installation of potentially intrusive software on the client computer to do so.

On the server, the full ROS\footnote{Robot Operating System, \url{https://ros.org/}.} environment, needed to run \emph{rViz}, \emph{rqt} (including the \emph{interactive\_script} plugin \citep{Witte.2019}), and the quadcopter simulation, was executed in a docker container\footnote{\url{https://www.docker.com/}. The container with the PLE is available on our github page at \url{https://github.com/sp-uulm/interactive_script/releases/tag/StudyPLE}.} accepting remote connections through \emph{vnc}. In order to reduce the complexity of the server infrastructure, only a single instance without scaling to support multiple concurrent users was used; therefore only one participant at a time could use the PLE.

\subsection{Programming tasks -- Quadcopter missions}
In the accompanying instruction PDF, users were asked to implement three different quadcopter missions which involved planning three flight trajectories (see~\ref{sec:instruction_pdf}). As we aimed to lower the initial barrier of starting to program, we added eight additional hints about programming in general. For instance, users were advised to divide one challenging task into easier sub-tasks to solve the missions. Furthermore, a short explanation about the different components of the interactive PLE was included. Depending on their experimental conditions, users were informed about the dynamic links between the code and the preview. Before beginning with the first mission users had to simply recreate a given example (see Figure~\ref{fig:studyoverview}). This example aimed at enabling users to get to know the interactive PLE and to the mapping aid between the code and the preview if available.

\subsection{Questionnaires}
In an online pre-questionnaire, participants were asked for their gender, age, educational level, the field of studies, and any prior experiences with programming and specific programming language. To gain further insights into their domain-specific prior knowledge, we developed eight questions about basic concepts of programming and relevant concepts for the programming missions. A pre-test included four knowledge questions to measure domain-specific prior knowledge: In addition to three basic definitions of the terms GUI, debugging, and logical operators, users also had to name a way point in the coordinate system. To capture a more in-depth understanding of programming concepts, another four questions were created (e.g. ‘\textit{Name one major difference between dynamically and statically typed programming languages.}’). To ensure good quality of measurement of prior knowledge, we analyzed the inter-rater reliability, which revealed very high consistency between the two ratings ($r = .993$, $p < .001$, $CI_{95\%} = .989 - .995$). 

To measure users‘ need for cognition, we used the German version of the short scale for need for cognition \citep{beissert2014}. Users had to rate four items (e.g. ‘\textit{I like my life to be full of tricky tasks that I have to solve.}’) on a seven point Likert-scale. In line with the published retest-reliability of $r = .78$, the scale was assumed to be reliable \citep{beissert2014}. In our sample, we determined a McDonald's omega of $\omega=.731$ ($CI_{95\%} = .640 - .823$) which was acceptable.

As both spatial ability and general logical ability are relevant to interact with the PLE as well as learning the relevant concepts, we included a measurement for fluid intelligence. We used the figural modul of the BEFKI 11+  including sixteen items \citep{schipolowski2017}. Dimitrov's scale reliability for the published sample is $\rho = .81$ \citep[total scale gf where the figural model is a part in;][]{schipolowski2017}.

For measuring learning outcome, we developed a post-test, which consisted of nine questions on the  three levels \cite[][ knowledge, comprehension, application]{Bloom.1956}. To measure the learning outcome on knowledge level, three multiple choice questions were developed (e.g. '\textit{Which statements about variables are correct? a) Only numbers and words can be assigned as values. b) A variable can be assigned multiple times. c) A variable can always be used instead of the value assigned to it. d) The value of a variable is unchangeable.}'). For the comprehension level, two open questions were included (e.g. '\textit{Name a basic difference between the pose block and the wait block.}'). The second comprehension question included three short programming sequences and their effects on the flying trajectory should be described. For application, we developed four questions. Two questions refereed to finding bugs in a given code. In the other two questions, users needed to map corresponding code and preview as well as describe the change in the code based on two different simulation results in the preview. In the present study, the questions of the post-test aimed to examine different aspects of the learning content. Different questions referred to different concepts and processes. Hence, we expected no high internal consistency of the different questions. To ensure that learning outcome was measured in a rigorous way, we analyzed the inter-rater reliability, which revealed a very high consistency between the two ratings ($r = .999$, $p < .001$, $CI_{95\%}  = .998 - .999$). 

Novices' cognitive load was measured using the differentiated cognitive load questionnaire \citep{klepsch.2017}. This questionnaire contained 2 items for intrinsic cognitive load (ICL), reflecting the complexity and element interactivity of the content. We used 3 items for assessing extraneous cognitive load (ECL), which reflects the load imposed by the design of the environment.  To test the resources novices invested when programming the three tasks and 3 items for germane cognitive load (GCL) were given. The respective items were subjectively rated on a 7-point Likert scale ranging from 1 (not at all) to 7 (completely). Results are reported as  z-scores for RQ4. The reliability was sufficient with $\alpha=.783$ ($CI_{95\%} = .670 - .861$) for ICL, McDonald's omega of $\omega=.876$ ($CI_{95\%} = .829- .923$) for ECL, and McDonald's omega of $\omega=.702$ ($CI_{95\%} = .595 - .808$) for GCL.

\subsection{Learning process data and program correctness }
To gain deeper insights into the approaches novices used to program the quadcopter mission, based on process data, traces of learning strategies were analyzed \citep{Hadwin.2007,Patwardhan.2017, WildKPSchiefeleU.1994}. In line with previous publication \cite[e.g.][]{Arendt.2021}, we defined certain indicators for different learning strategy traces.

As cognitive learning strategies, organizing and elaboration were considered. For organizing, it is assumed that novices worked with the preview to comprehend the structure of the code, but leave the code unchanged. For example, perspective changes were recorded here, or scaling of the preview (zoom-in(-out). To capture this, a threshold of 30 seconds was set in which no code changes took place, but the preview had to be changed. To assess elaboration in the PLE, the time during which novices were (actively) engaged in the task was considered.

For the metacognitive learning strategies, planning and monitoring were considered. The planning phase was measured as the time (after completion of a task) until the start of the (next) task. For monitoring, the number of started simulation runs in the preview was chosen as indicator.

Based on the manual savings and the screenshots at the end of each quadcopter mission, the novices solution was documented. The solution of each task was critically examined by two experts on the basis of a specific grading scheme. This allowed the objective assignment of points for the solution (program correctness) of each mission. In addition, the differentiated solution scheme allowed to identify typical errors in the tasks.

\subsection{Data preparation}
The online questionnaires were presented using the tool Unipark. We prepared and analyzed the data using R 4.1.0. Process data was analyzed both automatically and by using a specific labeling tool developed for this use case: consecutive unchanged screenshots were automatically filtered out to reduce redundant data. Timestamps of automatic saves and screenshots were then used to automatically classify user activity (e.g. changing the program, using the preview) and calculate metrics, such as time and duration of user activity. The start and end of programming exercises as well as usage of the mapping aids and simulation runs were manually labeled using a custom labeling tool to seek through and display screenshots and saved programs. A weighted mean of the prior knowledge test and the learning outcome scores were calculated for each user, which was based on the rating of two raters. For figural intelligence and need for cognition, the scores were calculated by using the published solution schemes \citep{beissert2014, schipolowski2017}. Furthermore, we checked for outliers and whether assumptions for parametric testing were met (normal distribution, homogeneity of variances, and covariances). To ensure high testing power for the analysis of our hypothesis and to avoid accumulation of alpha error, we used MANOVAs or MANCOVAs. Additionally, we used contrasts to analyze our hypotheses. By this, we had not to rely on the less powerful post-hoc testing \citep{Field.2013}.

\section{Results}
\label{sec:results}

\subsection{Descriptive Results and assumption testing}
The learners' domain-specific prior knowledge was medium, in accordance with the idea that our sample contains novices in programming (see Table~\ref{tab:descriptiveStatistics}). For figural intelligence, we found medium values in all four experimental conditions (see Table~\ref{tab:descriptiveStatistics}). To ensure that there were no systematic differences between experimental groups, we conducted a MANOVA that included prior knowledge, figural intelligence, and need for cognition. Considering learners’ preconditions, no significant differences could be found between the experimental groups ($F < 1$, $p >.454$). Furthermore, multivariate normal distribution was tested but not found for all critical variables for each experimental group. Conducting parametric MANCOVA is still assumed to be robust and favorable compared to non-parametric testing based on recommendations of \citep{Finch.2005}. Homogeneity of variance was supported based on the Levene-Test ($p > .062$). Homogeneity of covariances was checked using the Box-Test, based on which homogeneity was assumed ($p>.868$). As both, program correctness and learning outcome were claimed to be performance measures, a positive relationship between these two concepts was expected. In line with this, we found a strong, positive correlation $r=.71$ between program correctness and learning outcome. 

\begin{table}[htbp]
\centering
\caption{Means and standard deviations of the experimental groups on different variables}
\label{tab:descriptiveStatistics}
\begin{tabular}{llcccc}
\toprule
 & & No support & Highlights & Dynamic & Both mapping \\
 & &            &            & linking & aids \\
 & & $n = 19$ & $n = 22$ & $n = 21$ & $n = 20$ \\
 & & M (SD) & M (SD) & M (SD) & M (SD) \\
\cmidrule{3-6}
\multicolumn{2}{l}{\textit{Novices’ characteristics in \%}} & & & & \\
 & Prior knowledge             & 43.37 (19.92) & 37.50 (19.92) & 38.19 (18.60) & 45.99 (20.78) \\
 & Figural knowledge           & 55.40 (17.50) & 57.20 (18.10) & 58.80 (17.20) & 54.80 (10.80) \\
 & Need for cognition          & 66.35 (17.91) & 61.53 (19.28) & 64.29 (16.56) & 66.43 (15.26) \\
\multicolumn{2}{l}{\textit{Dependent Variables in \%}}       & & & & \\

\multicolumn{2}{l}{Program correctness}                       & & & & \\
 & Mission 1                   & 76.32 (19.50) & 79.55 (17.01) & 72.22 (15.21) & 90.00 (13.68) \\
 & Mission 2                   & 67.54 (33.09) & 64.39 (24.83) & 62.70 (23.51) & 85.00 (20.16) \\
 & Mission 3                   & 26.75 (39.14) & 27.27 (41.96) & 33.33 (39.44) & 54.17 (46.48) \\
 \multicolumn{2}{l}{Learning outcome}                         & & & & \\
 & Knowledge                   & 66.23 (18.94) & 62.88 (22.96) & 67.46 (20.40) & 71.67 (23.32) \\
 & Comprehension               & 29.39 (33.55) & 32.67 (23.77) & 41.27 (27.40) & 46.98 (25.55) \\
 & Application                 & 49.67 (31.63) & 42.61 (33.33) & 51.79 (26.01) & 66.88 (26.99) \\
\bottomrule
\end{tabular}
\end{table}

\subsection{Inferential testing} \label{IT}
We conducted a MANCOVA including the different experimental groups represented by the two factors highlighting and dynamic linking to analyze our hypotheses concerning main effect of the mapping aids. As covariates prior knowledge, figural intelligence and need for cognition were chosen. As dependent variables, we included program correctness of the three quadcopter missions (RQ1) and all three levels of learning outcome (knowledge, comprehension, application; RQ2). To test our hypotheses, we included simple contrasts for each mapping aid (highlights, dynamic linking). For analyzing the synergetic effect, we additionally included contrasts comparing the three experimental conditions with mapping aids.
To analyze our hypothesis concerning the relationship between learning strategy traces and performance, a bivariate correlation was utilized (RQ3 \& RQ4). Additionally, we focused on the collected process data. For gaining insights into the probability of an exemplary typical error we calculated a logistic regression (RQ5) and calculated z-scores to examine the novices' characteristics in the context of our sample (RQ6). 

\begin{table}[htbp]
	\centering
	\caption{Results of the between-subject effects of the MANCOVA with the different levels of learning outcome, program correctness, simple contrasts for element-to-element mapping aid (highlights), and relation-to-relation mapping aid (dynamic linking), with learners’ figural fluid intelligence, prior knowledge and need for cognition as covariates.
	}
	\label{tab:mancova}

\begin{tabular}{llccr}
\toprule

&                             & $F(1,75)$ & $p$     & $\eta^2_{\footnotesize\textit{partial}}$    \\ 
                             \cmidrule[0.4pt]{1-5}
\textit{Knowledge}

&Highlights                   & $0.01$    & $.940$  & $<.001$ \\ 
&Dynamic linking                       & $0.58$    & $.450$ & $.008$           \\ 
&Dynamic linking \textasteriskcentered{} highlights          & $0.44$    & $.507$  & $.006$            \\ 
&Prior knowledge              & $0.93$    & $.338$  & $.012$            \\ 
&Figural intelligence         & $49.0$    & $.486$ & $.006$            \\
&Need for cognition           & $0.03$    & $.872$ & $<.001$ \\ 
&          &         &       &  \\ 
\textit{Comprehension}

&Highlights                   & $1.26$    & $.265$  & $.017$            \\ 
&Dynamic linking                         & $5.61$    & $.020$  & $.070$            \\ 
&Dynamic linking \textasteriskcentered{} highlights            & $0.08$    & $.782$  & $.001$           \\ 
&Prior knowledge              & $0.10$    & $.752$ & $.001$           \\ 
&Figural intelligence         & $0.36$    & $.552$  & $.005$           \\ 
&Need for cognition           & $1.19$    & $.279$  & $.016$           \\ 
&          &         &       &  \\ 
\textit{Application}
&Highlights                   & $1.00$    & $.321$ & $.013$            \\ 
&Dynamic linking              & $4.08$    & $.047$  & $.052$            \\ 

&Dynamic linking \textasteriskcentered{} highlights            & $2.55$    & $.114$  & $.033$           \\ 
&Prior knowledge              & $0.97$    & $.327$  & $.013$            \\ 
&Figural intelligence         & $4.21$     & $.044$  & $.053$            \\
&Need for cognition           & $11.5$    & $.001$  & $.133$            \\ 
&          &         &       &  \\ 
\textit{Program correctness}
 &Highlights                   & $7.47$    & $.008$ & $.091$            \\ 
 \textit{mission 1}
&Dynamic linking                        & $0.23$    & $.636$ & $.003$            \\

&Dynamic linking \textasteriskcentered{} highlights            & $3.64$    & $.600$  & $.046$           \\ 
&Prior knowledge              & $0.00$    & $.994$  & $<.001$ \\
&Figural intelligence         & $0.44$    & $.511$  & $.006$            \\ 
&Need for cognition           & $3.72$    & $.058$  & $.047$            \\
&          &         &       &  \\ 
\textit{Program correctness}
&Highlights                   & $2.59$    & $.112$  & $.033$            \\ 
\textit{mission 2}
&Dynamic linking                       & $0.94$    & $.335$  & $.012$            \\

&Dynamic linking \textasteriskcentered{} highlights            & $3.83$    & $.054$  & $.049$           \\ 
&Prior knowledge              & $1.13$    & $.291$  & $.015$            \\
&Figural intelligence         & $0.41$    & $.525$  & $.005$            \\
&Need for cognition           & $6.93$    & $.010$  & $.085$            \\ 
&          &         &       &  \\ 
\textit{Program correctness} 
&Highlights                   & $1.34$    & $.252$  & $.017$            \\ 
\textit{mission 3}
&Dynamic linking                       & $1.91$    & $.171$  & $.025$            \\ 
&Dynamic linking \textasteriskcentered{} highlights            & $0.99$    & $.324$  & $.013$           \\ 
&Prior knowledge              & $2.70$    & $.104$  & $.035$            \\ 
&Figural intelligence         & $6.34$     & $.014$  & $.078$            \\ 
&Need for cognition  & $3.37$    & $.070$  & $.043$            \\
\bottomrule
\end{tabular}
\end{table}

\subsection{Effects on Program Correctness (RQ1)}
As both mapping aids outlined inter-relations between the different parts of the interactive PLE, we expected for both mapping aids a beneficial effect. 
As mentioned to analyze our hypotheses, we used the results of the conducted MANCOVA including the contrasts for each mapping aid, and the program correctness of all three quadcopter missions. 

\begin{figure}[H]
    \centering
    \includegraphics[width=\textwidth]{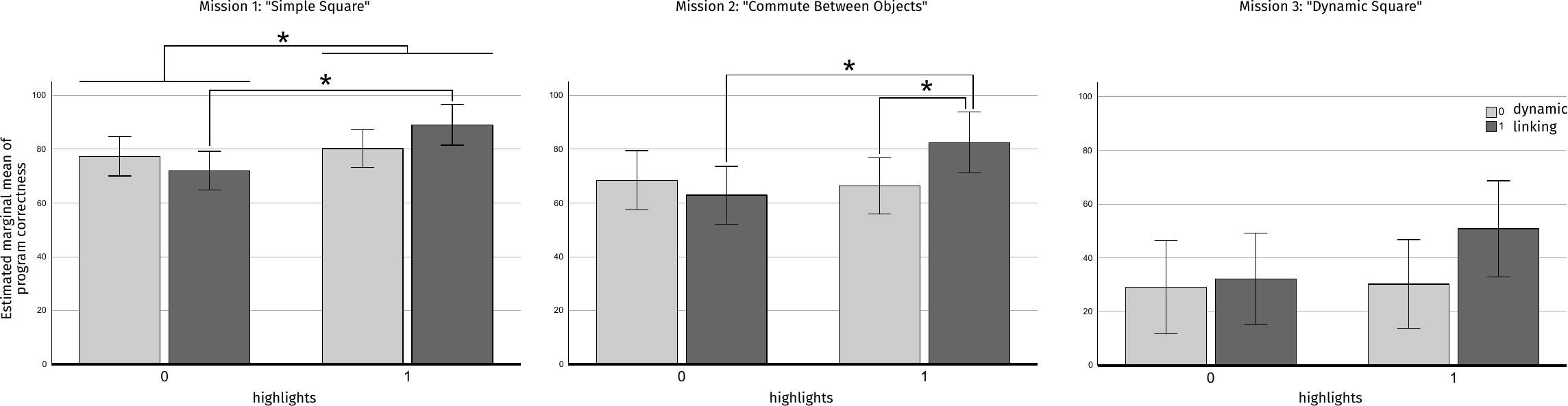}
    \caption{Means and standard error ($CI=95\%$) of the program correctness (in $\%$) of the three quadcopter missions depending on the experimental group.}
    \label{fig:rq1}
\end{figure}

We expected a beneficial effect of highlighting on program correctness (H1a). Based on the descriptive means of program correctness in all three tasks, the means were higher in experimental groups with highlighting in mission 1 and 2 (see Figure~\ref{fig:rq1}).

The expected significant, beneficial effect was only found in the first mission (mission 1: $p=.008$*, $\eta^2_{\footnotesize\textit{partial}}=.091$) but not for the following missions (mission 2: $p=.112$, $\eta^2_{\footnotesize\textit{partial}} =.033$; mission 3: $p=.252, \eta^2_{\footnotesize\textit{partial}}=.017$). Thus, our hypothesis was only partially supported by the data (H1a). 

Additionally, we expected a beneficial effect of dynamic linking for solving the programming tasks correctly (H1b). Based on descriptive means, no substantial difference was found in the first task between the groups with and without dynamic linking. Descriptively, higher means were given in the groups with dynamic linking compared to the groups without dynamic linking in the second and third task (see Figure~\ref{fig:rq1}).

We found no significant, beneficial effect for dynamic linking on solving the three tasks (mission 1: $p=.636$, $\eta^2_{\footnotesize\textit{partial}} <.003$; mission 2: $p=.355$, $\eta^2_{\footnotesize\textit{partial}} =.012$; mission 3: $p=.171$, $\eta^2_{\footnotesize\textit{partial}}=.025$). Based on these findings, our hypothesis was not supported (H1b).

As both mapping aids eased the programming of the quadcopter missions, we expected a synergetic effect of both mapping aids, compared to one mapping aid (H1c). In addition to the reported results of the MANCOVA (see Table~\ref{tab:mancova}), we analyzed the differences between the groups with mapping aids with respective contrasts, to gain insights into whether the combined (synergietic) mapping aid (dynamic linking \& highlighting) results in significant higher performance compared to the single mapping aid conditions (dynamic linking or highlighting). We found a significant beneficial synergetic effect for both mapping aids compared to dynamic linking for the first and second mission ($p_{\footnotesize\textit{mission1}}=.002$*, $p_{\footnotesize\textit{mission2}}=.015$*) but not for the third mission ($p_{\footnotesize\textit{mission3}}=.135$). Additionally, we found a beneficial, synergetic effect for both mapping aids compared to highlighting for second mission ($p_{\footnotesize\textit{mission2}}=.042$*) but not for the first or third mission ($p_{\footnotesize\textit{mission1}}=.095$, $p_{\footnotesize\textit{mission3}}=.099$). 

Hence, the hypothesis (H1c) was partially supported by the data.

\subsection{Effects of highlighting and dynamic linking on learning outcomes (RQ2)}

\begin{figure}
    \centering
    \includegraphics[width=0.8\textwidth]{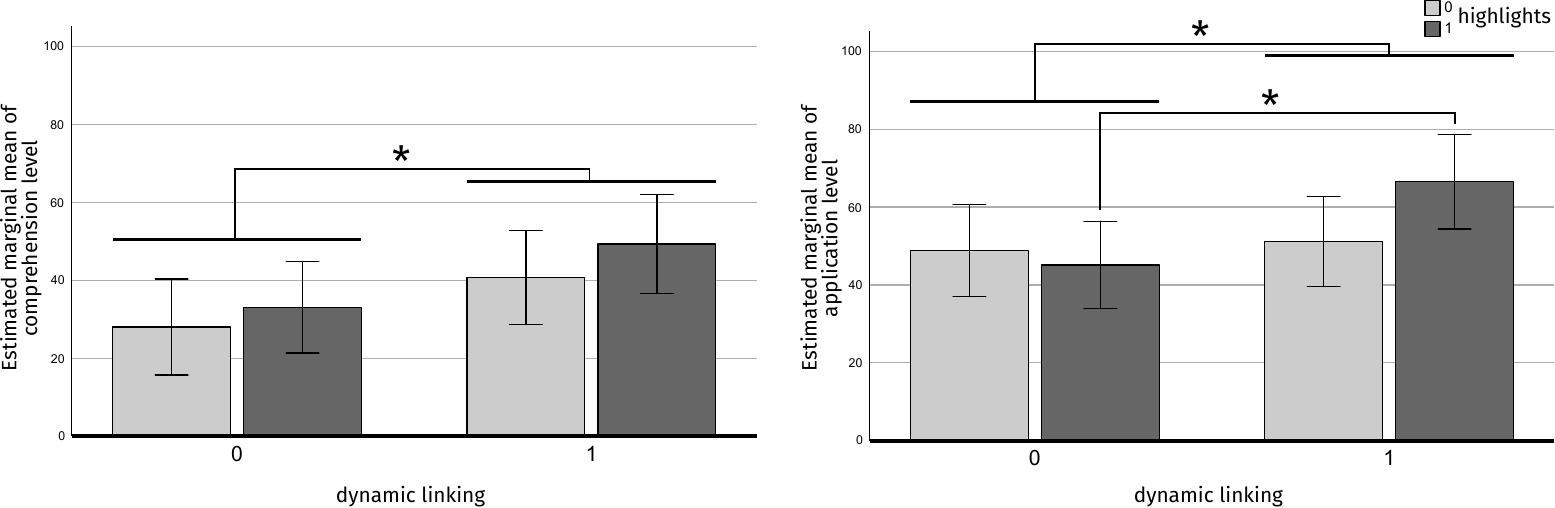}
    \caption{Effects of mapping aids on comprehension (left) and application level (right) under consideration of the covariates displaying means and standard errors ($CI=95\%)$).}
    \label{fig:rq5}
\end{figure}

We expected, that we would find stronger beneficial effects of highlighting on the comprehension and the application level of learning outcome compared to the knowledge level (H5a). Descriptively, we found higher means for comprehension and application level in the experimental group with highlighting compared to the control group while the knowledge level did merely differ (see Figure~\ref{fig:rq5}).
Based on our MANCOVA, we found no significant main effect of highlighting on the three levels of learning outcome ($p>.265$, see Table~\ref{tab:mancova}). 
Thus, our hypothesis was not supported by the data (H2a). 

In our next hypothesis, we expected the strongest beneficial effect of dynamic linking on comprehension and application level (H2b). Descriptively, in the groups with dynamic linking, the means of all levels of learning outcome were higher compared to the groups without dynamic linking. We found a significant main effect of dynamic linking on the comprehension level of learning outcome $F(1,75)=5.61$, $p=.020$*) with a medium effect size ($\eta^2_{\footnotesize\textit{partial}} =.070$) and a significant effect on application ($F(1,75)=4.08$, $p=.047$*, $\eta^2_{\footnotesize\textit{partial}} =.052$): There were no significant beneficial effects of dynamic linking on the knowledge ($p=.450$, $\eta^2_{\footnotesize\textit{partial}} =.008$). Hence, our hypothesis was supported by the given data as the strongest effect of dynamic linking was found for the comprehension and application level (H2b).

Comparing the different mapping aids for their beneficial effect on different learning outcome (H2c), the descriptive pattern implied stronger beneficial effects of both mapping aids compared to a single one (see Figure~\ref{fig:rq5}). For testing hypothesis H1c, we used a contrast analysis to complement the MANCOVA results and to allow insights into the specific differences between the groups with mapping aids. 

In our hypothesis, we assumed a synergetic effect of both mapping aids compared a single mapping aid on the different levels of learning outcome (H2c). Based on our MANCOVA contrast analysis, we found a significant beneficial synergetic effect for both mapping aids compared to highlighting on application level ($p_{\footnotesize\textit{application}}=.013$), only ($p_{\footnotesize\textit{knowledge}}=.317$, $p_{\footnotesize\textit{comprehension}}=.068$, see Figure~\ref{fig:rq5}).

In contrast, comparing both mapping aids and dynamic linking ($p_{\footnotesize\textit{knowledge}}=.678$, $p_{\footnotesize\textit{comprehension}}$ $=.335$, $p_{\footnotesize\textit{application}}$=.074) did not reveal significant differences. Hence, our hypothesis (H2c) was only partially supported by the data.

\subsection{Correlation of Learning Strategy Traces \& Program Correctness (RQ3)}

Having a closer look at the learning behavior while programming the quadcopter mission, traces of cognitive (organizing, elaborating) and meta-cognitive (planning, monitoring) learning strategies were analyzed. We expected the traces to be positively correlated to overall program correctness. 

\begin{figure}[H]
    \centering
    \includegraphics[width=0.7\textwidth]{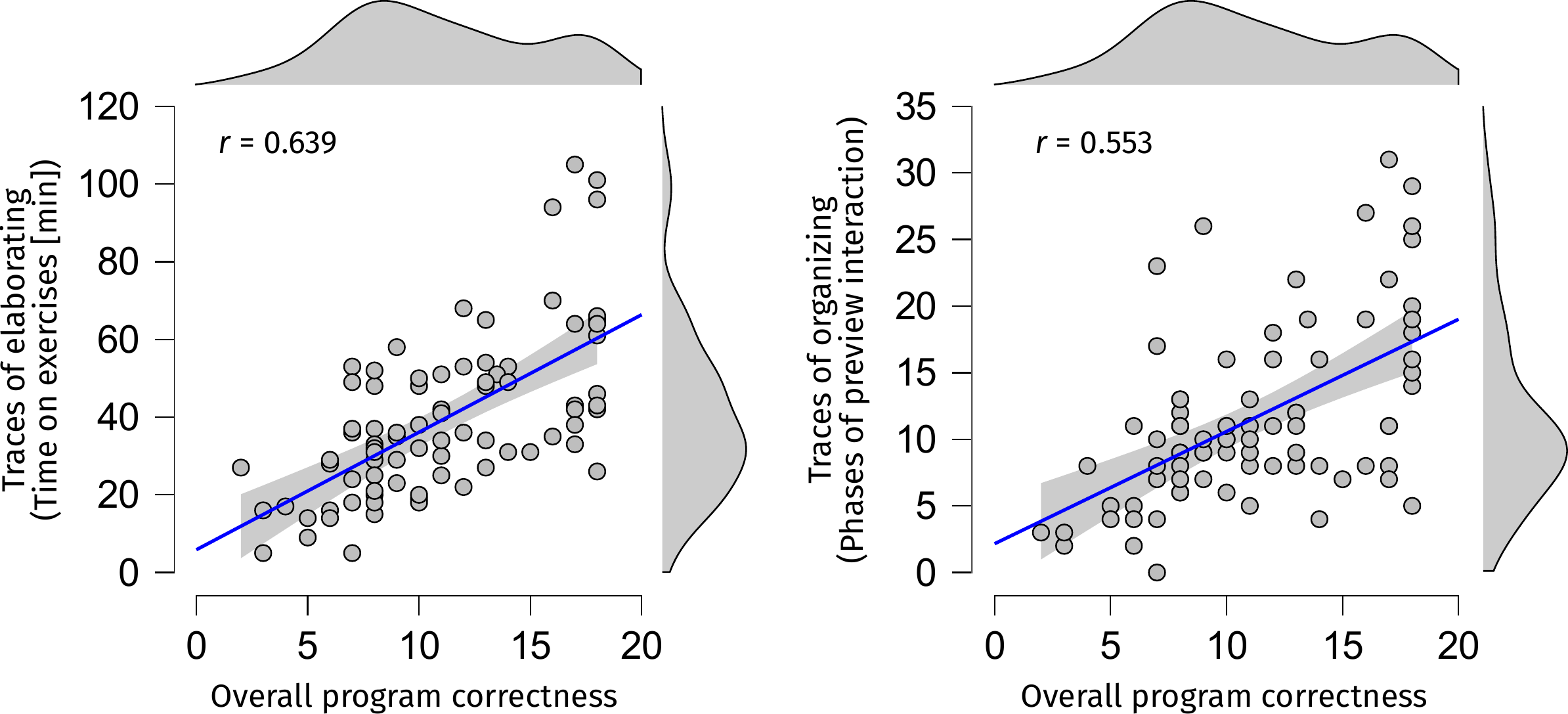}
    \caption{Scatter plot displaying the relationship between traces of cognitive learning strategies and overall program correctness.}
    \label{fig:rq3}
\end{figure}

First, we analyzed the traces of the cognitive learning strategies. Based on scatter plot inspection, we found the expected pattern of organizing (indicated by interacting only with the graphical preview) and program correctness. We found a medium, positive correlation ($r=.553$, $p<.001$***, see Figure~\ref{fig:rq3}). 

Thus, our hypothesis (H3a) was supported. Inspecting the descriptive pattern of elaborating (trace indicated by overall time in the PLE), a positive relation with program correctness was expected. In line with our hypothesis (H3b), we revealed a strong, positive correlation between elaboration and program correctness ($r=.639$ $p<.001$***).
Having a look at the traces of meta-cognitive learning strategies, based on the visual inspection, a positive relationship between planning (indicated by time before starting the tasks) and program correctness was not plausible. In line with this, no significant correlation was found ($r=-.163$, $p=.214$) and the hypothesis (H3c) was not supported by the data. For monitoring (indicated by starting the simulation in the preview), we found the pattern fitting to our hypothesized relation with program correctness. However, we found no significant correlation between monitoring and program correctness ($r=.234$, $p=.056$; H3d).

\subsection{Relation of learning traces and learning outcome (RQ4)}
\begin{figure}[H]
    \centering
    \includegraphics[width=0.7\textwidth]{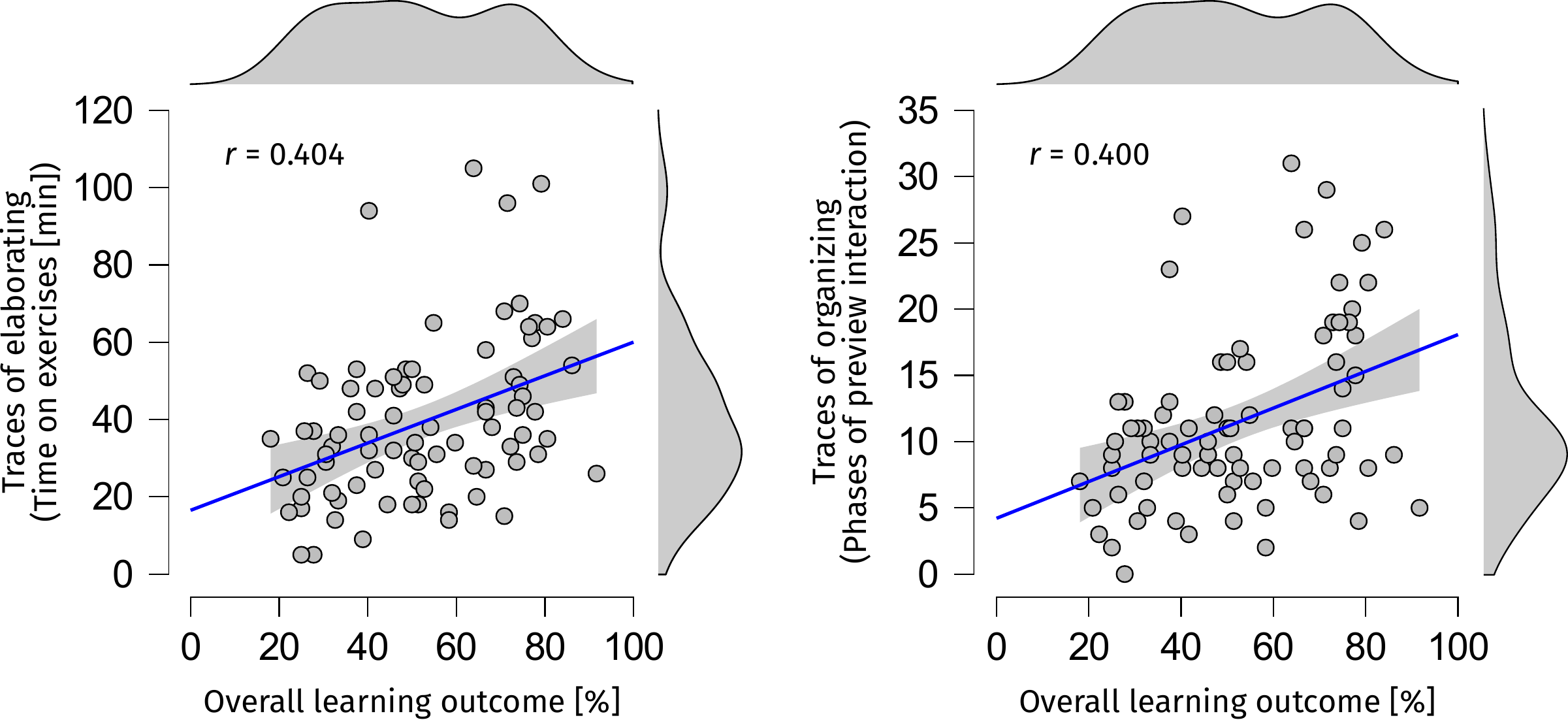}
    \caption{Scatter plot displaying the relationship between traces of cognitive learning strategies and overall learning outcome.}
    \label{fig:rq6}
\end{figure}

We expected traces of cognitive and meta-cognitive learning strategies to be positively related with overall learning outcome. In line with the positive descriptive trend, we found a medium, positive correlation between traces of organizing and overall learning outcome ($r=.400$, $p<.001$***, see Figure~\ref{fig:rq6}). Thus, our hypothesis (H4a) was supported. 

Inspecting the descriptive pattern of elaborating, a positive relation with learning outcome was expected. In line with this the findings related to our hypothesis (H4b) revealed a medium, positive correlation between elaboration and overall learning outcome ($r=.404$, $p<.001$***, see Figure~\ref{fig:rq6} ).
Having a look at the traces of meta-cognitive learning strategies, again (see RQ3) a positive relationship between planning and program correctness was not plausible. In line with this, no significant correlation was found planning ($r_{\footnotesize\textit{monitoring}}=.169$, $p=.170$) nor monitoring ($r_{\footnotesize\textit{planning}}=-.013$, $p=.924$). Thus, these hypotheses were not supported by our findings (H4c \& H4d).

\subsection{Typical Errors (RQ5)}\label{res:rq5}

In order to receive further insights into the novices handling of the programming missions, we analyzed process data. This method allowed further insight into the aspects where mistakes were typically made. Since we were dealing with a very large sample, process data from the first mission were analyzed for sources of error. These resulted from the solution scheme. The first mission was evaluated on the basis of these criteria. To achieve the maximum score of 6 points, the following criteria had to be met: Creation of four different waypoints; these waypoints form a square; angles of the quadcopters were manipulated at least once; angles of the quadcopters were correctly adjusted as desired in the task; the quadcopter flies; each "moveTo" is followed by a "wait" or "sleep".

In the first mission, setting the angle of the drones proved to be a typical source of error. Descriptively, it was found that about 57\% ($n=47$) of the participants did not set the angle correctly and the distribution seemed to differ between the experimental groups.

Therefore, this typical error was analyzed in more detail as an example. To test whether the mapping aids offered had an effect on the performance in setting the correct angle, we performed a logistic regression including the covariates (see Table~\ref{tab:log_reg}).  Based on our regression, participants with both mapping aids had a 5.18-times higher chance ($p=.020$*) of having set the angle correctly in the first task compared to the control group. Hence, our hypothesis was supported (H5) for this exemplary typical error. For further insights into other typical errors (see~\ref{sec:typical_errors}). 

\begin{table}[htbp]
	\centering
	\caption{Coefficients of the logistic regression predicting the chance of correct angle in mission 1 considering the effect of experimental groups (1-3) against the control group (4) with figural intelligence, need for cognition and prior knowledge as covariates}
	\label{tab:log_reg}
	{
		\begin{tabular}{lrrrrrrrr} 
			\toprule
			\multicolumn{1}{c}{} & \multicolumn{1}{c}{} & \multicolumn{1}{c}{} & \multicolumn{1}{c}{} & \multicolumn{1}{c}{} & \multicolumn{1}{c}{} & \multicolumn{3}{c}{Wald Test} \\
			\cline{7-9}
			 & \textit{Estimate} & \textit{SE} & \textit{Odds Ratio} & $z$ & \textit{Wald $\chi^2$} & \textit{df} & $p$  \\
			\cmidrule[0.4pt]{1-9}
			(Intercept) & $-2.387$ & $1.427$ & $0.092$ & $-1.673$ & $2.799$ & $1$ & $.094$  \\
			Figural intelligence & $0.418$ & $1.509$ & $1.519$ & $0.277$ & $0.077$ & $1$ & $.782$  \\
			Need for cognition & $0.013$ & $0.015$ & $1.013$ & $0.859$ & $0.738$ & $1$ & $.390$  \\
			Prior knowledge & $0.011$ & $0.013$ & $1.011$ & $0.843$ & $0.710$ & $1$ & $.399$  \\
			Dyn. linking \textasteriskcentered{} highlights (1) & $1.645$ & $0.706$ & $5.181$ & $2.330$ & $5.429$ & $1$ & $.020$  \\
			Dynamic linking (2) & $-0.077$ & $0.704$ & $0.926$ & $-0.109$ & $0.012$ & $1$ & $.913$  \\
			Highlights (3) & $0.537$ & $0.675$ & $1.711$ & $0.796$ & $0.634$ & $1$ & $.426$  \\
			\bottomrule
		\end{tabular}
	}
\end{table}

\subsection{Analysis of the profiles of successful novices (RQ6)}

\begin{figure}[H]
    \centering
    \includegraphics[width=0.6\textwidth]{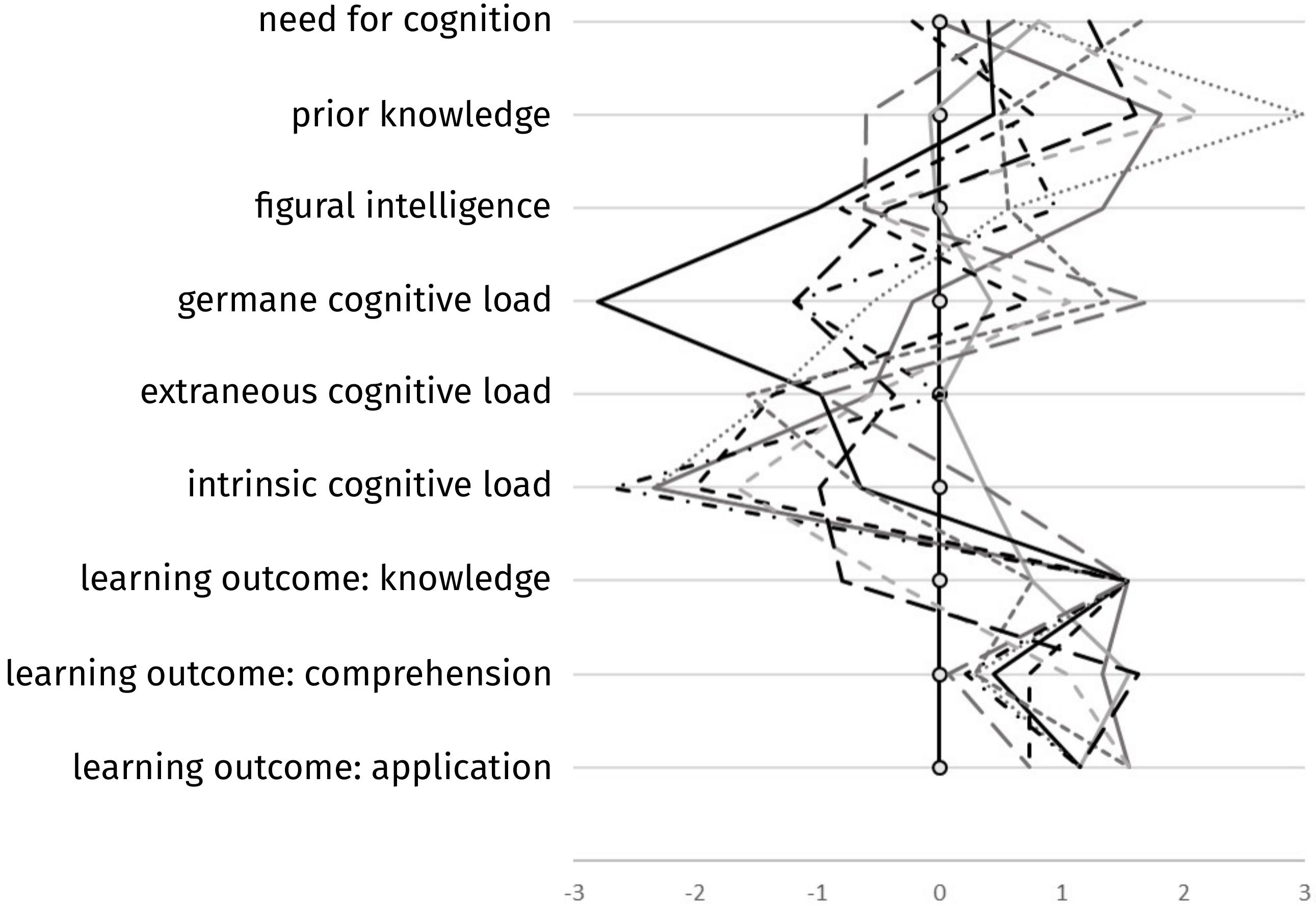}
    \caption{Z-scores of different variables characterizing the top ten users based on program correctness.}
    \label{fig:rq4}
\end{figure}

By analyzing the characteristics of the 10 participants that reached all points in the programming exercises, we gain further insights into the profiles of successful novices (see Figure~\ref{fig:rq4}): the normalized scores for the measured characteristics of each of these participants are plotted and connected to detect possible common tendencies or deciding characteristics. Of the 10 selected participants, 5 had both help features available, 2 had only highlights, 3 had neither dynamic linking nor highlights, and none had only dynamic linking enabled.

In the subgroup, tendencies can be described for the different criteria (H6): Successful users in programming the three missions tended to have higher scores on the characteristic need for cognition, mostly above-average prior knowledge, and figural intelligence. Looking at the cognitive load of the learners in this subgroup, no clear pattern emerges with regard to the germane cognitive load, but extraneous cognitive load and intrinsic cognitive load tend to be lower compared to the overall mean. The users in this group rated the usability of the PLE above average. The top 10 users in terms of program correctness were also in the successful group in the learning test, particularly in the areas of comprehension and application. Thus, there is a certain profile of successful users and as described earlier, programm correctness and learning outcomes were related.  However, for example, in prior knowledge and figural intelligence and knowledge learning outcome no clear pattern can be derived. Since this last research question is more of an exploitative question, it can be concluded that some of the results (e.g. learning outcome) correspond to our hypothesized expectations, but no clear and consistent profile of successful novices can be described. Hence, our hypothesis was partially supported by the findings (H6).

\section{Discussion}
\label{sec:discussion}

The effect of mapping aids were considered based on a multi-method assessment approach. By this, we gained further insights into different aspects of potential benefits from the respective mapping aids. 

\subsection{Mapping aids and program correctness (RQ1)}
We expected a beneficial effect of the two mapping aids on program correctness. In contrast to our initial expectation, only highlighting increased performance for programming the first mission but no significant, beneficial effects were found for dynamic linking. Additionally, the assumed synergistic effect was only found when comparing both mapping aids against dynamic linking (first and second mission) and highlighting (second mission).

Despite the fact that a positive effect of dynamic linking and combined mapping aids compared to single-help conditions is theoretically plausible based on cognitive processes, previous empirical studies describe similar findings as in our study \cite[e.g.][]{Rey.2011}.  With respect to the underlying cognitive processes, plausible reasons for the lack of a positive effect of dynamic linking can be described in addition to the specific application-related conclusions: Additional load on working memory due to the presentation and use of the aids might occur and result in a lower performance \citep{Lowe.1996, sweller2020}. Reflecting on this, \cite{Rey.2011} concluded based on theoretical inferences and the heterogeneous empirical findings that dynamic linking is not per se advantageous or disadvantageous, but that the specific implementation also plays an important role. Additionally, \cite{Seal.2010} described, that interactive elements might support users depending on the given task and emphasized that mapping aids might cause additional strain while learning. The three quadcopter missions had different levels of difficulty: While mission 1 only required the extension of the given code example, the subsequent missions contained further code elements and calculations that were not explicitly included in the example. The question now is whether certain aids or their combination are particularly advantageous for different partial solutions of the given missions. This question was examined in RQ5 (see Section~\ref{dic:rq5}). 
In addition, the question arises as to what effect the mapping aids have at the learning level and whether, for example, they particularly support the comprehension and application of what has been learned. This was examined in more detail in the next research question.

\subsection{Mapping aids and learning outcome (RQ2)}
Based on our hypotheses, we expected a beneficial effect of the two mapping aids on higher levels of learning outcome. Beneficial effects were found for dynamic linking on comprehension and application level of learning outcome but not for highlighting. Comparing single aids to the combined condition revealed beneficial effects for adding dynamic linking on the application level. 

Based on these findings, different conclusions can be made: (i) the knowledge level of learning outcome was not affected by the mapping aids or their combination. This was in line with our expectation as these mapping aids facilitated the integration of elements and processes in the PLE and not in particular the sub-semantic processing of the different components \citep{Patwardhan.2017, fries2021}. (ii) Beneficial effects on learning outcomes were found for dynamic linking for both comprehension and application while highlighting did not significantly increase the learning outcome. This was not in line with our initial expectations. However, having a closer look at previous findings, a heterogeneous pattern for effects of mapping aids can be described. In contrast to the described advantages, there are also studies \citep[e.g.][]{vanderMeij.2006} that do not describe the beneficial effects of mapping aids. Based on theoretical considerations, plausible explanations for these findings exist: When using highlights, novices are passively pointed to connections and no longer actively reflect on them, and thus no deeper learning occurs \citep{Ainsworth.1999, Seufert.2006}. 
Providing highlights in combination with a very specific task could, for example, encourage the use of means-end strategies \citep{erhel2019}.

Since finding e.g. two connected elements and then the interaction of these is particularly central for the level of understanding, this could be a possible explanation why highlights were rather less helpful in answering these questions.
In order to understand this finding more deeply, future studies should include further measurements for the development of mental models in order to better differentiate further sub-processes of information processing (for details see \citep{Vogt2021diss}. This will provide more precise insights into the effects of highlights on cognitive processes. 

For dynamic linking, novices where hinted to related processes, but to benefit in terms of learning outcome, they still needed to reflect the relation between the respective components where as highlighting already made the relevant relation very explicit \citep{Gentner.1983}. (iii) This is also reflected by the finding concerning the synergistic effect of both mapping aids compared against the single aid conditions on learning outcome: novices were stimulated by adding dynamic linking to process the learning content more deeply, which was reflected in synergistic effects on the application level when comparing both mapping aids. 
\subsection{Traces of learning strategies and performance (RQ3 \& RQ4)}

When analyzing the relationship between the learning strategies and their relationship to performance, it was found that only the cognitive strategies traces were positively related to performance (program correctness and learning outcome). In contrast, these relationships with performance could not be found for the metacognitive strategies traces.

Reflecting critically on these findings, the following explanations can be found: Although the selected indicators or traces can be assigned to specific learning strategies, there are other possible indicators as well \citep{roll2015}. Based on our findings, metacognitive strategy traces were not related to performance. In our study, planning was traced by the time before the onset of programming each mission. However, it cannot be ruled out that novices also planned during the tasks. Hence, our finding reflect one general challenge when dealing with process data and deducing intentions or strategies based on behavioral traces \citep{malmberg2017}. 

In addition, the instructions for the missions were already transparently divided into substeps or subproblems to accommodate novices. Hence, a part of the structuring process and planning process was already transparent. Therefore, it is plausible that planning in this specific case is not strongly related to performance, as it was already facilitated by the task instruction. Hence, the operationalization of the metacognitive strategy trace needs to be reconsidered in future studies as well as the metacognitive strategy use during performing the respective programming tasks.

In contrast, traces of monitoring showed a descriptive trend indicating a positive correlation with program correctness, which was not significant. Again, the question is to what extent the monitoring already took place when the novices looked at the preview without running the simulation and whether further or better traces for the monitoring process can be found here, which would allow even deeper and more differentiated insights into the application of metacognitive learning strategies.

\subsection{Avoiding typical errors (RQ5)}\label{dic:rq5}
In order to gain further insight into the solution of the programming missions, typical errors in mission 1 were analyzed. We focused our analysis on the first mission as the high rate of successful novices lead to few but common mistakes that were well suited for detailed analysis. The lower success rate in, e.g., mission 3 leads to a high number of very diverse observed problems -- as well as many incomplete solutions -- that mask interesting, typical errors. The exemplary typical error was that novices did not set the correct angle of the quadcopter or did not adjust it at all. The analysis showed that novices with both aids had a significantly higher chance of setting the angle correctly.

Reflecting in detail on the role of the mapping aid in solving this partial problem, the findings are in line with expectations based on the theoretical assumptions. For instance, dynamic linking offers a direct manipulation of the angles. Additionally, highlighting should be helpful for the orientation and assignment of the respective way points. When combining both mapping aids, novices should be supported in the solution of the angle setting in the first quadcopter mission. Hence, this exemplary partial problem was chosen to examine whether the odds of error changed depending on the available mapping aids. Typical problems for correct angle stetting were, that novices either did not change the angle at all, or were not able to deduce the correct values for the required angles for each way point. In the condition with both mapping aids, they had the opportunity to adapt the 3D preview directly to set the angles correctly, without the dynamic linking, they had to calculate the angle. Without the highlights they had to search for the respective way point without assistance to map the code and the way point in the preview.

In the future, based on the available empirical studies, guidelines could be developed, which would allow the type of mapping assistance to be deduced in the context of the specific (sub)problem and thus enable targeted assistance in the individual (sub)tasks \citep{Seal.2010}. To gain further insight, in addition to analyzing the effect of aid, the relationship between certain strategies and performance was also examined in more detail. 

\subsection{Successful patterns and individual requirements for success (RQ6)}
In the last research question, the novices' profiles of relevant characteristics were examined in detail. By including some of the characteristics as a control variable, the influence of the characteristics on the performance was controlled to avoid bias on the performance measures. Nevertheless, it seems worthwhile to take a closer look at the profiles of the 10 most successful novices. As expected based on the high correlation between program correctness and learning outcome, the 10 most successful novices (based on program correctness) were also above average based in learning outcome. 

However, the cognitive load during the tasks showed a very heterogeneous pattern of findings. The cognitive resources invested (germane cognitive load) varied greatly. Thus, it cannot be assumed that success in the missions depended significantly on the cognitive resources provided, only. The intrinsic cognitive load, which is based on task difficulty and element interactivity, also showed a high degree of variation. Here, a plausible assumption would be that successful novices subjectively rate the tasks as easier and are then correspondingly more successful \citep[for further details of the cognitive load theory see:][]{de2017}. This assumption was not supported by the present findings. Another plausible assumption would be that novices with more prior knowledge or better skills (figural intelligence, which correlates substantially with intelligence) are more successful \citep{kell2018}. However, these two characteristics are also clearly scattered in the group of the most successful novices. 

One characteristic was outstanding: need for cognition. This concept focuses on the extent to which people like to deal with complex (thinking) tasks and the extent to which they actively seek out such tasks in everyday life itself. This characteristic also seems to be relevant for the successful programming of the missions or problem solving in general \citep[e.g.][]{Coutinho.2005,rudolph2018}. Therefore, it also represents a possible starting point for supporting novices in their entry into programming. If people are encouraged to take more enjoyment and be more active in dealing with complex tasks (even in their free time) and positive incentive systems (e.g. serious games) are created, then learning programming languages could be promoted in this way \citep{rudolph2018}.

\subsection{Limitations of the study and threats to validity}

Due to the pandemic situation, the PLE was used as an online tool. This setting went along with some limitations. In a controlled laboratory study, the investigator could have ensured that the available aids were used (at least once). Despite the explicit hint to use the mapping aids on the base of the initial example, not all subjects complied with this request. Furthermore, it could not be ensured that the novices did not use other (online) support to answer the questions or to solve the programming tasks. To limit this problem, the tasks were very specific and the simple copying of code examples was not possible due to the Blocky implementation used. Nevertheless, this could be controlled for more effectively in a follow-up laboratory study. 

Another limitation of the online PLE was that the novices participated on their own computers. For example, participants had different screen sizes, interactions (touch pad/mouse), and individual settings, including a possibly slow internet connection, may have also contributed to potential bias. In order to keep possible biases as small as possible, the section of the screen that was viewed was kept constant, regardless of the screen size. Nevertheless, it is conceivable that the screen size had an influence on the usability or readability of the provided PDF with the instructions. 

In our study, we collected a convenience sample. The main goal in recruiting was to find people with little programming experience. Based on the prior knowledge data, the sampling method used succeeded in finding novice programmers. Nevertheless, the sample contained mainly students. Psychology students were over represented due to their lower programming experience in comparison to, e.g., engineering students. In this sample, it can be observed that, for example, the figural intelligence was slightly above average. It would therefore be interesting to see whether the present findings can be replicated with other more heterogeneous samples and what role further learner characteristics (e.g. age, working memory capacity etc.) play for the positive effect of the mapping aids on performance and their use in the PLE. Some of the reported effect sizes were significantly smaller than the initially assumed effect size on which our a priori power analysis was based. In the future, larger samples could be obtained outside the pandemic period, as it was extremely difficult to find subjects during this time.

In the present study, the initial barrier to start programming and implement a solution was kept as low as possible. To this end, some hints and helpful information to ease the problem solving aspect of programming have been provided in the task description. With the help of these hints, novices could divide the tasks into sub-steps and to understand the structure of the task and the flight trajectory to be created.
This led to a simplification of the programming tasks in order to make it possible for novices to solve different tasks in less time and focus on the implementation and program maintenance aspects. The library of available program blocks and the domain specific language used was tailored to fit the programming tasks -- this reduces the time to introduce all necessary concepts and language constructs. Nevertheless, the environment and language can be easily extended and adapted to real world applications without adding much complexity. The programming tasks are therefore rather artificial and only reminiscent of real world scenarios, which would be more complex, time-consuming, and require a longer training.

\section{Conclusion}
\label{sec:conclusion}

Domain experts operating cyber-physical systems often lack in depth programming knowledge. Therefore, the design of the programming (and learning) environment as well as support features to help acquire basic knowledge of the system are crucial to successfully program the CPS. Source location tracking provides valuable information to automatically link elements of a live preview to corresponding code elements. Mapping aids that support novice learners in solving programming tasks can be implemented using this information.

In the present study, we examined the effects of a \emph{highlighting} feature that highlights relevant locations in the code when clicking on an element in the preview, and a \emph{dynamic linking} feature, that allowed direct manipulation of the code through the preview, on program correctness and the learning outcome of novice learners. Participants were tasked to implement different quadcopter missions in a block based PLE with a 3D live preview and simulation.

While for program correctness highlighting was beneficial, dynamic linking showed more positive effects on deeper learning. The combination of both aids in comparison to single aids also revealed positive effects. The analysis of typical errors shows that especially the combination of both helps increased the chance of a correct solution. Traces of strategies when working on the missions can already anticipate success in programming and learning. No consistent profile for the best performing novices  (based on their characteristics) was found but higher programm correctness was strongly related to learning outcome.

\paragraph*{Future Work}

Technological solutions could also be used to gain further insights into the use of the PLE. In the present study, we have already used many different parameters to investigate the effects of the mapping aids on performance and behavioral patterns. In addition, psychophysiological measurements (e.g., eye-tracking) could provide further insight into the associated cognitive processes. Other formats such as video or audio recordings could be used as possible sources of information. For example, learners could be actively encouraged to verbalize their thoughts and solution processes (think aloud), possibly providing further insights into the strategies used. This could reduce speculation and uncertainty while classifying and interpreting the intent of the participant and thereby improve the transparency of behavioral patterns and cognitive processes.

In our study, we found a positive relationship between traces of cognitive learning strategies (organizing, elaborating) and performance based on process data. In future studies, indicators of these traces could be tracked online through learning analytic approaches and used as a live feedback tool to provide helpful inputs (e.g., cognitive prompts) that could trigger the use of learning strategies when needed \citep[e.g.][]{zumbach2020}. In this way, individual and adaptive PLEs could be implemented, and if sufficient training data are available, artificial intelligence approaches could also be utilized to provide feedback or appropriate supporting elements. For this, the analysis of further typical errors could be helpful. If adaptation through artificial intelligence is conceptualized at different levels (micro and macro) and over longer periods of time, further insights could be gained into whether, for example, these factors are also important for finding effects of metacognitive strategy traces. 

Overall, our study offers only a limited temporal insight. It would also be interesting to observe novices in the process of learning to program for a longer period of time and to investigate whether the effect of the aids and their use changes over time and to what extent the learning effects found are long-term.

In future studies, the research question could be extended and the concept of the planning process could be explored in more detail, for example in connection with the concept of computational thinking \citep[e.g.][]{hodhod2014}. In this way, the present findings could be integrated into a larger context.

The display concept of the study could also be expanded to different settings: we developed, both, an augmented and a virtual reality PLE to investigate how mapping aids assist novices when the 3D preview is presented as an augmented reality element in the real laboratory, or whether presenting our PLE in virtual reality interferes with the positive effects of our mapping aids. 

We plan to implement and evaluate more advanced mapping aids and other live programming features that use and showcase the rich information SLT enabled PLEs can provide. These include improvement of proposed code changes through context information, live evaluation and live debugging tools that can explain program behavior using provenance traces, and test case generation or automatic program repair using SLT enabled assertions.

More adaptive PLEs could be developed in future studies that additionally detect and reinforce positive patterns with prompts and specifically support novices. SLT is a promising and adequate way to provide the necessary information to implement those novel help features.

\newpage
\paragraph*{Acknowledgement} This work was partially supported by the German Research Foundation (DFG): 435878599, 453895475, and the German Federal Ministry of Education and Research (BMBF): 16DHB2205.


\newpage
\bibliography{bibfile}

\newpage

\appendix

\section{Programming Exercises}
\label{sec:instruction_pdf}

The following pages in this section contain a translated version of the complete instructions and exercises given to the participants preserving the original formatting and layout. Parts that were only included in the instructions for specific experimental conditions are marked accordingly. 



# Introduction

Today you are supposed to program quadrocopter missions using a special programming environment for novice programmers. Please read the following pages carefully before you start solving the tasks.

> **hint**: Create a relaxed environment! It is recommended to use a large screen and a mouse and to avoid problems while working on the tasks.

## Programming and learning environment

The main window is divided into several sections: In the programming window pieces from the (expandable) library can be dragged. Below, a preview of the program output or possible errors is displayed. To the right of it a live preview of the program is displayed.

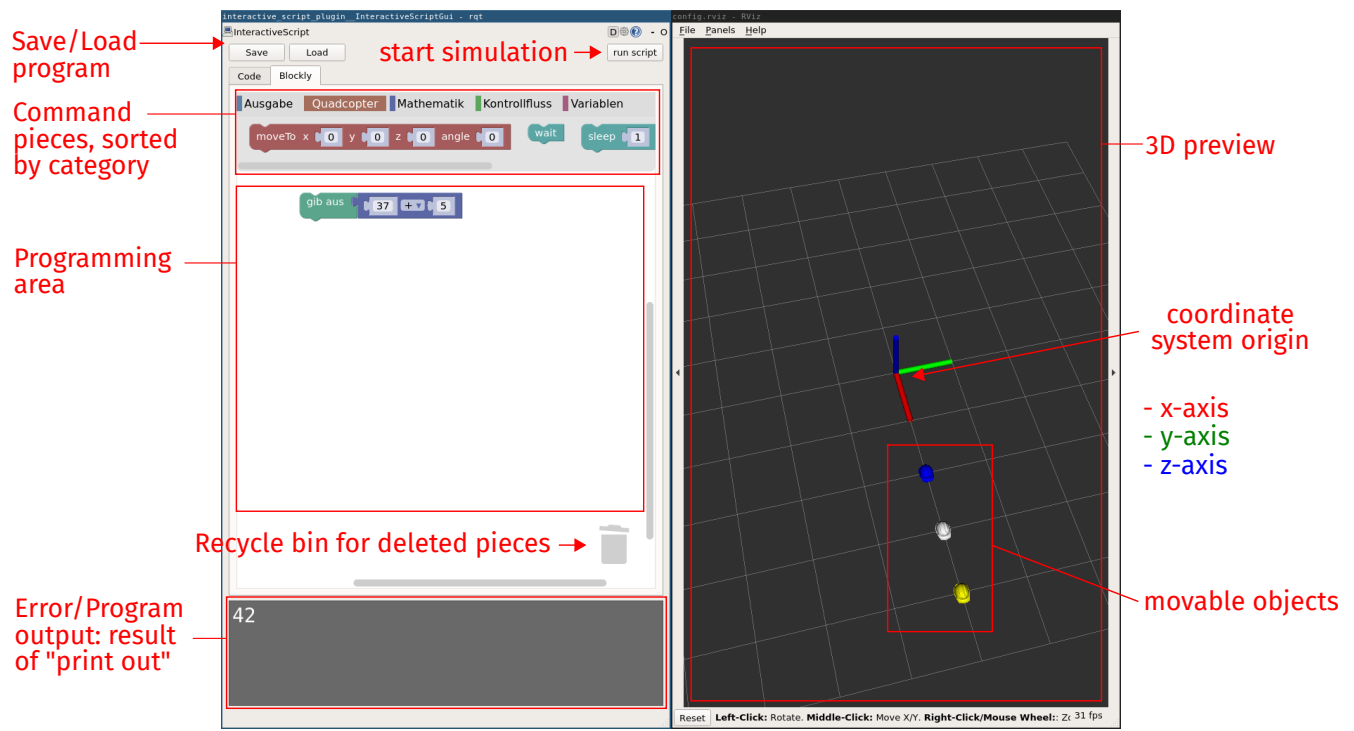

The pieces can be placed anywhere within the programming area, but it is recommended to form a coherent block that is processed from top to bottom.
Pieces can be copied (duplicated) by right-clicking on them.
Deleted pieces end up in the recycle bin. By clicking on the recycle bin, pieces can be dragged back to the program area.

## 3D preview

The 3D view shows a live preview of the current program. The view can be moved, rotated and zoomed with the mouse. The view can be rotated with the left mouse button, zoomed with the scroll wheel and panned around with the middle mouse button (click with the mouse wheel).
Feel free to try it out!
Waypoints are displayed as small white quadrocopters, the path between them as a straight line. If the line is red, the quadrocopter may not have enough time to reach the destination (for example, because the next waypoint is given directly without waiting pause), if it is white, the destination is reached.
In addition, the virtual space contains 3 colored objects that can be freely moved and rotated on the floor with the mouse.

## Interactive preview

*Note: This following section was only available in the respective experimental conditions.*

*For conditions with highlighting:*
Clicking on a waypoint will highlight in the programming window the values used to calculate the position of the waypoint. Changing one of the marked fields will move the waypoint, all other values have no influence on the selected waypoint.

*For conditions with dynamic linking:*
With the arrows a waypoint can be moved directly. It will automatically adjust the program so that the waypoint is located at the desired position and the changes are marked. However, there may be other ways to get the desired result or the change may have other, additional effects.

## Important programming commands

| | |
|---|---|
| 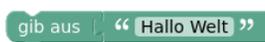 | Writes the value of the right part to the console (lower part of the window) |
| 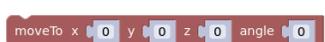 | Determine new positions on x, y, z (in m) and rotation (in °, around the z-axis) |
| 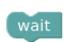 | Wait until the last set target position has been reached |
| 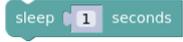 | Wait for 1 second |
| 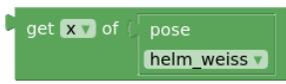 | Returns the current position and rotation of an object. Object can be 'hardhat_yellow', 'hardhat_white' or 'hardhat_blue'. The returned pose can be accessed with 'get ... of ...' to get a single value. |

## Coordinate System

Positions are always given in meters relative to the origin. This origin is located in the center of the grid. The grid lines form squares of 1x1m and mark the floor of the room. The x-axis is marked in red, the y-axis in green and the z-axis (altitude) in blue. The rotation around the z-axis is measured counterclockwise; $0°$ correspond to the direction of the x-axis.

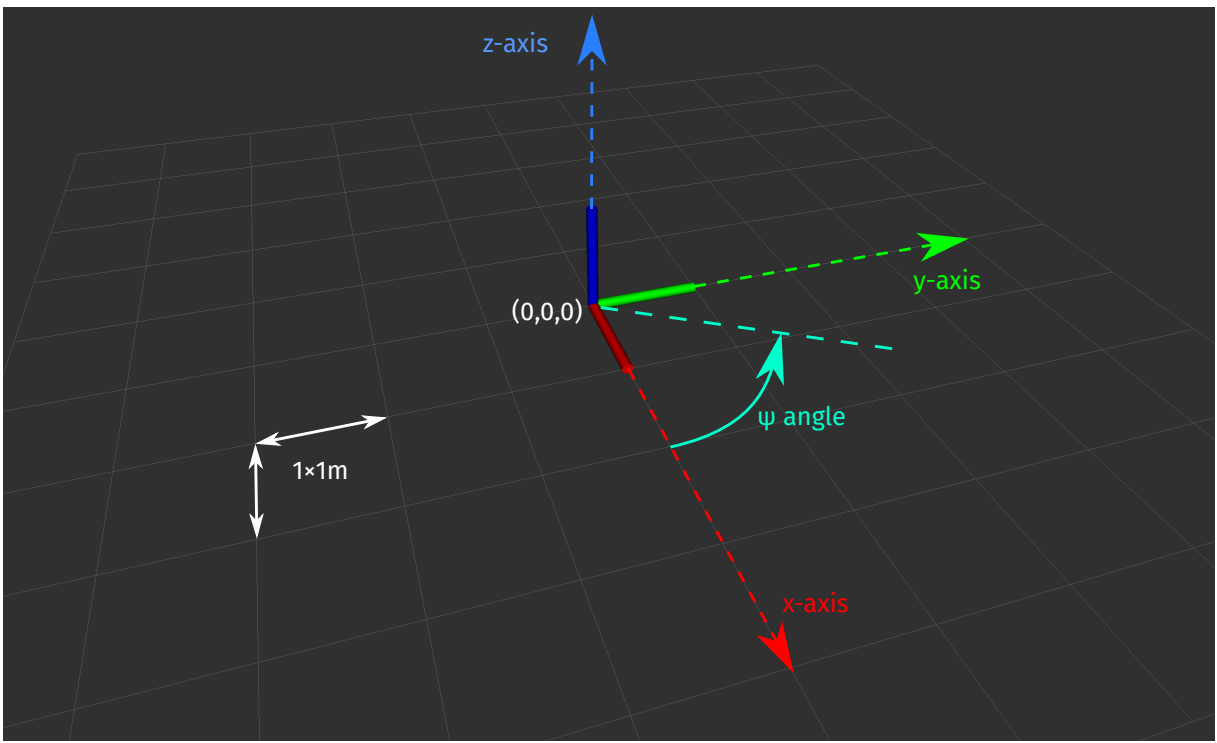

# 8 hints for solving the problems

1. **Programming is an iterative process**

   It is difficult to find the correct solution directly in the first attempt. Instead, try to create an incomplete solution first and improve it step by step. For example, if you are not sure about the position of the waypoints, it can help to place them somewhere first and then move them to the correct position using the preview.

2. **Experimentation is encouraged**

   Simulated quadcopters are cheap and can't break. It is explicitly recommended to try things even if you don't know exactly what will happen. The preview shows the expected program progress at any time. Additionally, by clicking on the 'run script' button, the program can also be simulated.

3. **Divide and conquer!**

   If the overall problem seems too complicated, try to identify and solve sub-problems. If these are still too complicated, break the problems down further.

4. **Display intermediate results**

   Any intermediate results can be displayed at any time with a 'print out' block. Instead of speculating whether a value is calculated incorrectly, it can simply be output and checked.

5. **Name intermediate results**

   Use variables to store intermediate results and to use them multiple times. You can also give them a name: don't be too shy to name the calculated side length of a square 'side length'.

6. **Take notes**

   A small drawing or note can be of great help. You don't have to solve everything on the computer.

7. **Use the preview**

   The preview is the perfect tool to check programs for errors. It's much faster to figure out where the point (2,1,3) is in space if you just set a waypoint to that position and look it up.

8. **Pay attention to hint boxes**

   The tasks contain hint boxes in many places that give useful clues to the solution. It doesn't hurt to read these hints several times if you feel stuck.

# Tutorial example [10min]

The quadcopter is supposed to fly up to a height of 2m above the origin and land again.

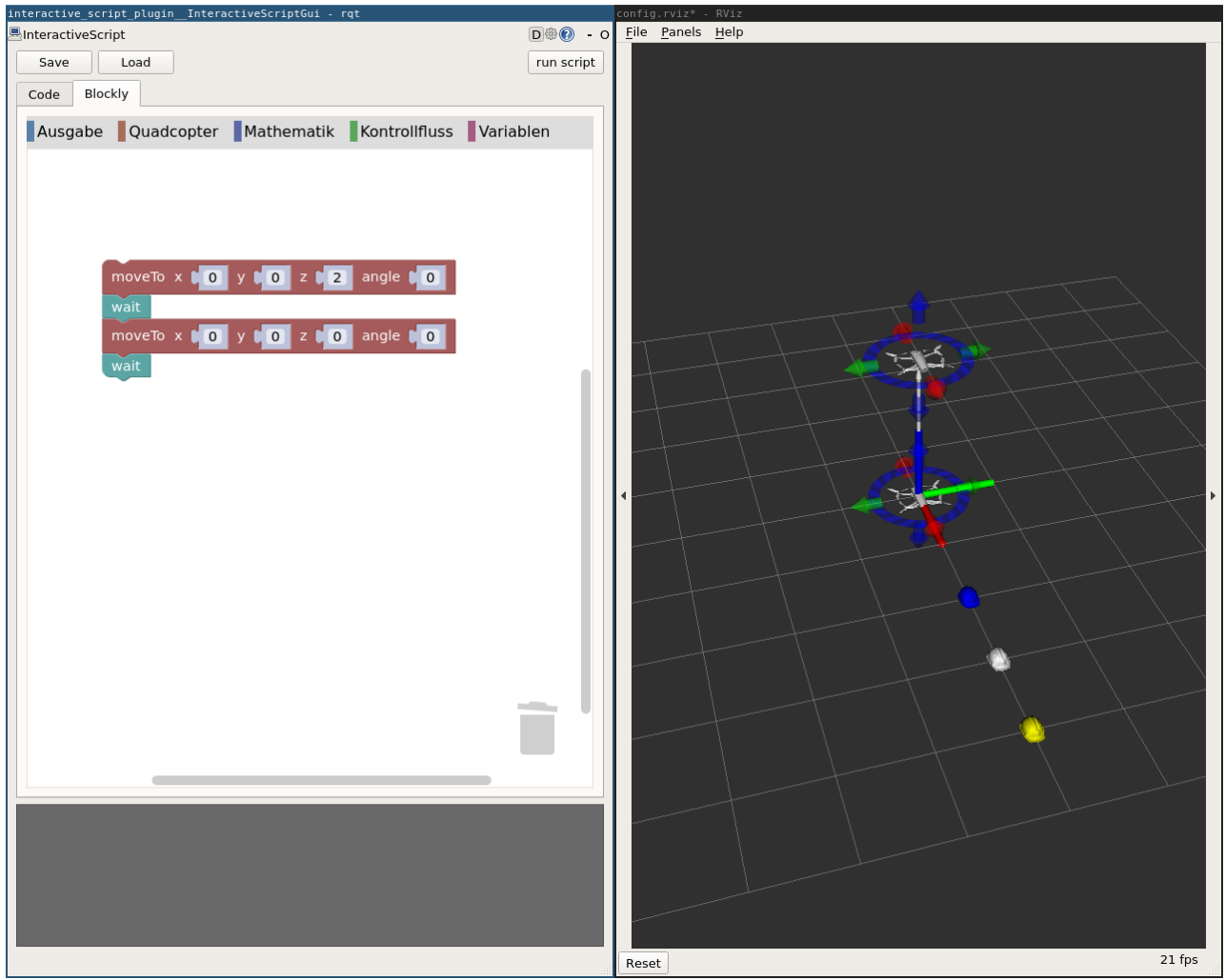

**TODO:**

☐ Replicate the given program.

☐ Look at the preview and make sure the waypoint is at 2m height.

☐ Run the simulation by clicking on the 'run script' button and compare the simulation with the preview.

☐ Try moving the waypoints in the x-direction and experiment with more waypoints or omitting the wait blocks and how that affects the simulation.

☐ When you are done click on 'Save' to mark the configuration as a solution and save it (a filename is automatically chosen and there is no feedback! The load function always automatically loads the last save state).

# Mission 1: Simple Square [10min]

The quadrocopter is supposed to take off, fly over the four vertices of a square and then land at the coordinate origin (0,0,0). To do this, familiarize yourself with the coordinate system and determine the four corner points of any square in flight space.

> hint: It is much easier to align the square to the axes of the coordinate system.

Create a program that makes the quadcopter fly over the four corner points one after the other. You already know the necessary commands from the previous example.

> hint: Note that after each new waypoint, it is necessary to wait so that the quadcopter also has the opportunity to reach the target!

Rotate the quadrocopter at each corner point so that it is headed towards the center of the square.

> hint:
> *Note: Only available in the respective experimental conditions.*
>
> *For conditions with highlighting:*
> You can click on waypoints to mark the corresponding places in the code.
>
> *For conditions with dynamic linking:*
> Waypoints in the preview can be moved directly.

**TODO:**

☐ Determine the corner points of the square.

☐ Create a program to fly off the square.

☐ Rotate the quadrocopter to the center at each corner point.

☐ Click on 'Save' to save the program.

# Mission 2: Commute Between Objects [15min]

Until now, the waypoints were fixed, but now the quadcopter is supposed to fly back and forth between (and above) the positions the white and blue helmet. This allows the quadcopter's flight trajectory to be changed during flight.

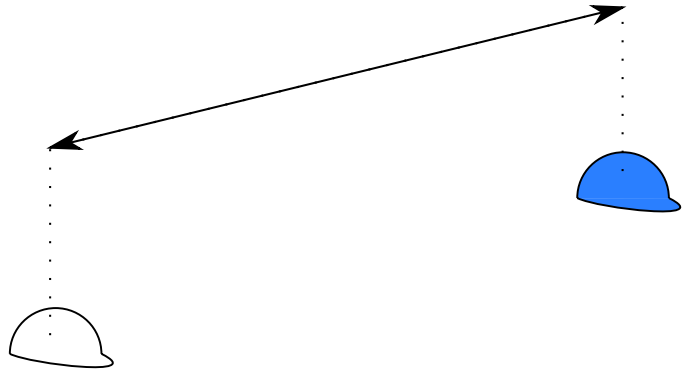

> **hint**
>
> The pose() function is special because it returns a result (the pose of the object at the time of the call). This result contains several components (x, y, z, angle) which can be accessed by the 'get ... of ...' block from the quadrocopter area (e.g. get x of pose('hardhat_blue') for the x-coordinate of the blue helmet). As an alternative, the result can also be cached and the individual components can be accessed later, i.e.:
>
> ![blocks: setze blau auf pose helm_blau; moveTo x 0 y get y of blau z 2 angle 0]
>
> To create a variable, select 'Create Variable...' and give it a name. Matching program parts will be created to assign a value to the variable and use the value. Any number of variables can be created as long as they have a unique name without special characters.

> **hint**
>
> Remember that you can output values textually in the gray console area with the 'print out' block. This is useful, for example, to display and check intermediate results.

> **hint**
> In order to avoid writing out repeatedly executed commands each time, a loop can be used. The commands within the loop are executed as often as specified in the loop header, e.g.
>
> 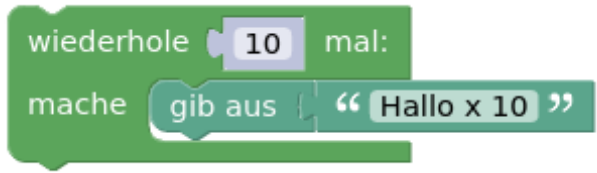

> **hint**
> The position of the helmet cannot be used directly, of course, because it is lying on the ground, so the z-coordinate should either be fixed or placed a bit above the helmet.

**TODO:**

- ☐ Write a program that determines the pose of the helmets.
- ☐ Calculate the waypoints of the quadcopter from the pose of the helmets.
- ☐ Write a loop so that the waypoints are flown to again and again in turn.
- ☐ Click on 'Save' to save the program.

## Mission 3: Dynamic Square [20min]

Now a little challenge: again a square is supposed to be flown off, but the length of its sides is supposed to be defined by the distance between the white and blue hardhats. For example, if the blue and white helmets are 1.63m apart, the length of an edge of the square should also be 1.63m.

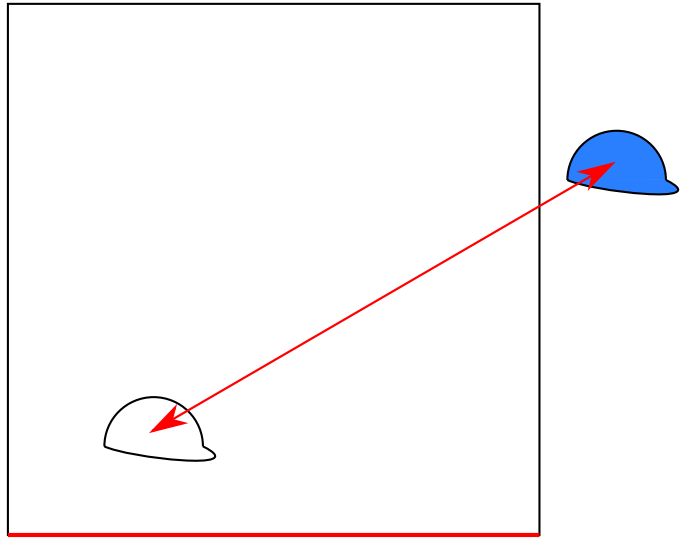

hint  One point of the square can be fixed, the other corner points have to be calculated based on the measured distance.

hint  The distance between two points $A$ and $B$ in space is $\sqrt{(x_A - x_B)^2 + (y_A - y_B)^2 + (z_A - z_B)^2}$. Corresponding blocks for the mathematical operations (e.g. square root) can be found in the math section.

hint  It is a good idea to assign the distance to a variable first in order to be able to use it several times.

hint  Try to solve the two problems (calculating distance and making the edge length of the square variable) separately: the vertices can be calculated depending on a variable that is assigned a fixed value as a test. Only when you are confident that the correct vertices are calculated for different distances, you should try to calculate the distance between the helmets and use the calculated value.

**TODO:**

☐ Write a program that calculates the vertices of a square of variable side length.



\newpage
\section{Typical errors while programming the three missions}
\label{sec:typical_errors}

\begin{figure}[htbp]
    \centering
    \includegraphics[width=0.7\textwidth]{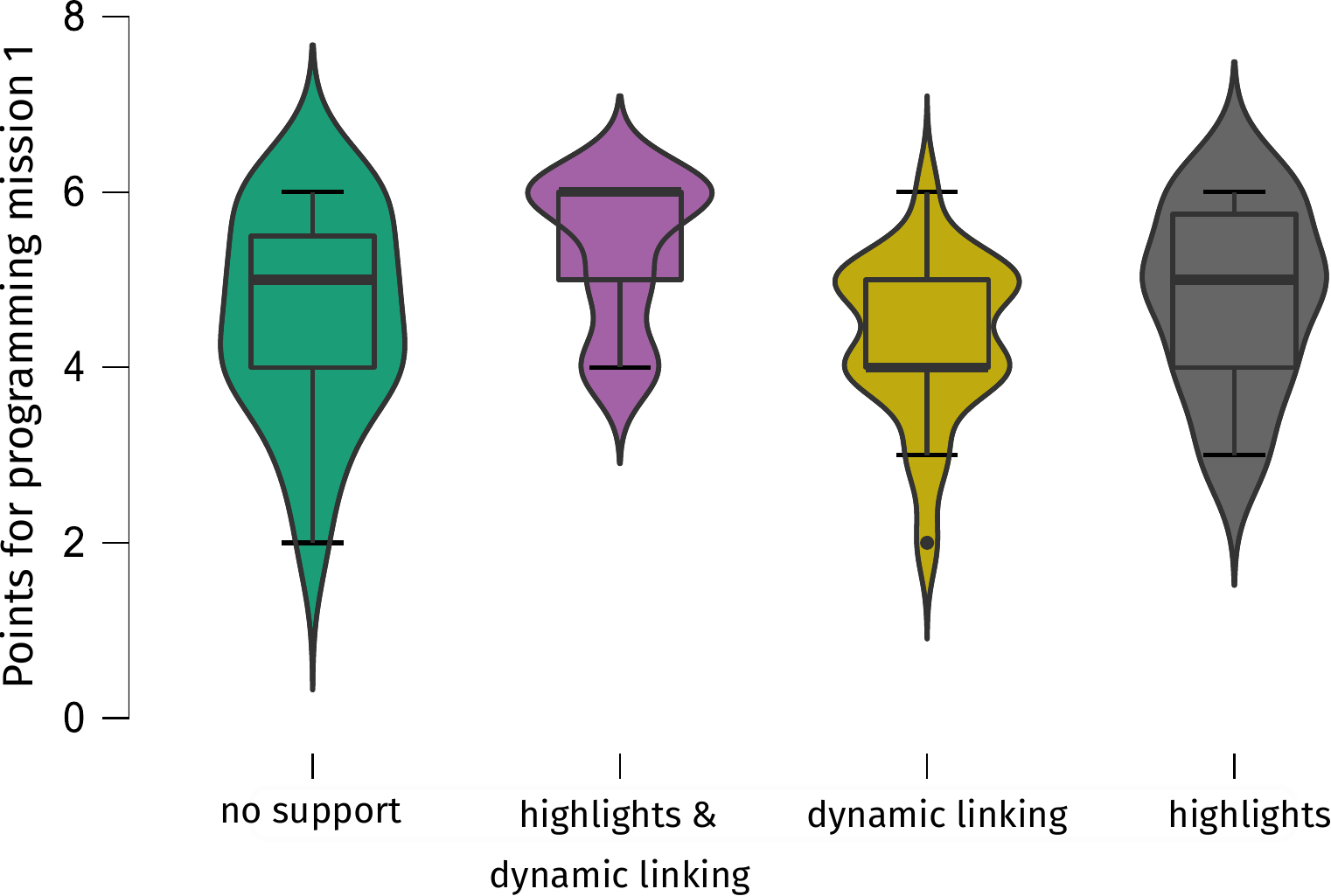}
    \caption{Violin plot displaying the point distribution for the first mission.}
    \label{fig:rq6a}
\end{figure}

In the following, we provide more insights into typical errors while programming the first mission. Based on our evaluation scheme, we analyzed individual errors that occurred. In general, the following errors were considered in the evaluation scheme:

\begin{itemize}
    \item quadcopter is flying ($z>0$)
    \item creating 4 waypoints
    \item waypoints result in a square
    \item at least 1x change of an angle
    \item correct angle setting
    \item each \textit{moveTo} is followed by a \textit{wait} or \textit{sleep} block
\end{itemize}
Overall, novices could achieve 6 point maximum. Only 8 of 82 scored 3 points or below in the first mission (see Figure~\ref{fig:rq6a}). 

Based on the violin plot displaying the descriptive pattern of the points in the first mission for each experimental group, differences can be observed between the four experimental groups (see Figure~\ref{fig:rq6a}).

Descriptively, most novices scoring 3 points or lower were found in the control group, followed by the group with dynamic linking as mapping aid. By analyzing the process data, including the automatically recorded screenshots during programming the mission, typical errors have been determined by two independent raters. Around 80~\% of the novices scored with 4 points or higher. Mostly, errors resulting in lower scores occurred based on three of the six evaluation criteria: the drone does not fly ($z = 0$),  the angles of the waypoints were not manipulated at all, and the angles were altered but the angle setting was not correct. The last aspect is described as an exemplary, typical error in the results section (see Section~\ref{res:rq5}). The other two typical errors are described in more detail in the following sections including a descriptive overview along with inferential testing using logistic regression approaches.

\newpage
\subsection{Quadcopters didn't fly}
\begin{figure}[htbp]
    \centering
    \includegraphics[width=0.9\textwidth]{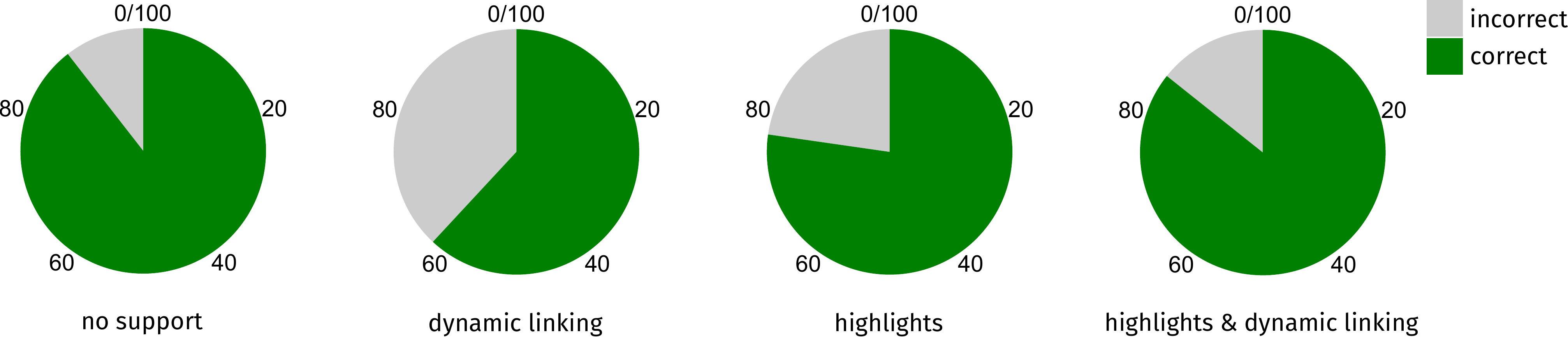}
    \caption{Pie charts displaying the proportion of drones that didn't fly (incorrect) and the flying drones (correct) for each experimental group.}
    \label{fig:rq6a_F}
\end{figure}

We analyzed the typical error not of altering the z-coordinate resulting in a quadcopter that does not fly. Descriptively, novices in the group with dynamic linking and/or highlights were less likely to alter the z-coordinate (see Figure~\ref{fig:rq6a_F}). The lowest descriptive probability of not altering the z-coordinate was found for the experimental group with dynamic linking.

To determine, whether this trend represents a significant effect, we conducted a logistic regression considering respective covariates. Our analysis revealed, that the dynamic linking group had a significant lower probability to adjust the z-coordinate compared to the control group (see Table~\ref{tab:rq6z}). This finding might support the idea described in the theory and previous literature, that the beneficial effect of mapping aids is also dependent of the considered (sub)task.

\begin{table}[h]
	\centering
	\caption{Coefficients}
	\label{tab:rq6z}
	{
	    \small
		\begin{tabular}{lrrrrrrr}
			\toprule
			\multicolumn{1}{c}{} & \multicolumn{1}{c}{} & \multicolumn{1}{c}{} & \multicolumn{1}{c}{} & \multicolumn{1}{c}{} & \multicolumn{3}{c}{Wald Test} \\
			\cline{6-8}
			 & Estimate & Std. Error & Odds Ratio & z & Wald Stat. & df & p  \\
			\cmidrule[0.4pt]{1-8}
			(Intercept) & $2.343$ & $1.650$ & $10.417$ & $1.420$ & $2.016$ & $1$ & $0.156$  \\
			Dyn. linking \textasteriskcentered{} highlights & $-0.376$ & $0.982$ & $0.687$ & $-0.383$ & $0.147$ & $1$ & $0.702$  \\
			Dynamic linking & $-1.790$ & $0.891$ & $0.167$ & $-2.010$ & $4.039$ & $1$ & $0.044$  \\
			Highlights & $-1.038$ & $0.918$ & $0.354$ & $-1.130$ & $1.278$ & $1$ & $0.258$  \\
			Figural intelligence & $1.303$ & $1.773$ & $3.682$ & $0.735$ & $0.540$ & $1$ & $0.462$  \\
			Need for cognition & $-0.007$ & $0.018$ & $0.993$ & $-0.383$ & $0.147$ & $1$ & $0.702$  \\
			Prior knowledge & $-0.010$ & $0.016$ & $0.990$ & $-0.646$ & $0.418$ & $1$ & $0.518$  \\
			\bottomrule
		\end{tabular}
	}
\end{table}

\newpage
\subsection{No angle adjustment}
\begin{figure}[htbp]
    \centering
    \includegraphics[width=0.8\textwidth]{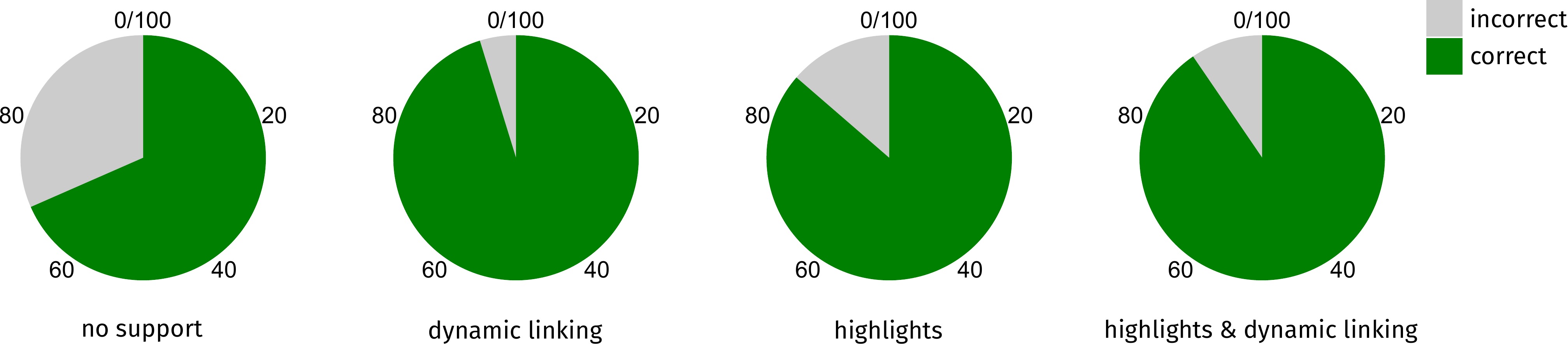}
    \caption{Pie charts displaying the proportion of no adaption of the angles (incorrect) and at least once adapted angles of the drones (correct) for each experimental group.}
    \label{fig:rq6a_AA}
\end{figure}

Next to the described angle setting error in the results section (see Section~\ref{res:rq5}), we additionally analyzed whether the provided mapping aids had an impact on the odds of changing the angle at least once. Descriptively, in the control group without mapping aids, novices were more likely to not adjust the angle at all (see Figure~\ref{fig:rq6a_AA}). 

We tested this typical error for significance by using a logistic regression considering different covariates: Significant effects of the mapping aids for both mapping aids and the dynamic linking were found (see Table~\ref{tab:rq6wa}). Novices in these groups were more likely to adapt the angle at least once compared to the control group, when the respective covariates were considered.

\begin{table}[h]
	\centering
	\caption{Coefficients of the logistic regression predicting the chance of at least one angle adaption in mission 1 considering the effect of experimental groups (1-3) against the control group (4)}
	\label{tab:rq6wa}
	{
	    \small
		\begin{tabular}{lrrrrrrr}
			\toprule
			\multicolumn{1}{c}{} & \multicolumn{1}{c}{} & \multicolumn{1}{c}{} & \multicolumn{1}{c}{} & \multicolumn{1}{c}{} & \multicolumn{3}{c}{Wald Test} \\
			\cline{6-8}
			 & Estimate & Std. Error & Odds Ratio & z & Wald Stat. & df & p  \\
			\cmidrule[0.4pt]{1-8}
			(Intercept) & $-1.581$ & $1.768$ & $0.206$ & $-0.894$ & $0.800$ & $1$ & $0.371$  \\
		Dyn. linking \textasteriskcentered{} highlights  & $2.218$ & $1.152$ & $9.189$ & $1.925$ & $3.707$ & $1$ & $0.054$  \\
			Dynamic linking & $2.351$ & $1.168$ & $10.495$ & $2.013$ & $4.050$ & $1$ & $0.044$  \\
			Highlights & $1.281$ & $0.840$ & $3.601$ & $1.525$ & $2.327$ & $1$ & $0.127$  \\
		Figural intelligence & $0.842$ & $2.120$ & $2.320$ & $0.397$ & $0.158$ & $1$ & $0.691$  \\
			Need for cognition & $0.026$ & $0.021$ & $1.026$ & $1.249$ & $1.560$ & $1$ & $0.212$  \\
			Prior knowledge & $0.005$ & $0.020$ & $1.005$ & $0.260$ & $0.067$ & $1$ & $0.795$  \\
			\bottomrule
		\end{tabular}
	}
\end{table}

\newpage
\paragraph*{Ethics Statement}
The study was carried out in accordance with the Declaration of Helsinki. Ethical review and approval was not required for the study on human participants in accordance with the local legislation and institutional requirements. The participants provided their written informed consent to participate in this study. Data was used pseudonymously and participants were aware that they had the chance to withdraw their data at any point of the study.

\paragraph*{Author Contributions - CRediT (Contributor Roles Taxonomy)}
TW and AV developed the initial design of the study (Conceptualization) which was feed backed by TS and MT (Supervision). AV and TW developed the missions and the study specific questions (learning outcome, prior knowledge etc.; Methodology). TW implemented the PLE and managed the technical infrastructure (Software; Visualization). AV led the data collection for the study (Investigation). AV and TW analyzed and interpreted the data (Formal analysis). AV and TW drafted the work (Writing – Original Draft), which was revised critically by MT and TS (Writing - Review \& Editing). All authors provided approval of the final submitted version of the manuscript and agree to be accountable for all aspects of the work in ensuring that questions related to the accuracy or integrity of any part of the work are appropriately investigated and resolved.

\end{document}